\documentclass[aps,superscriptaddress,amsmath,amssymb,twocolumn,amsfonts,floatfix,prl,longbibliography,10pt]{revtex4-2}
\usepackage{graphicx}
\usepackage{epstopdf}
\usepackage[T1]{fontenc}
\usepackage[latin9]{inputenc}
\usepackage{amsbsy}
\usepackage{gensymb}

\usepackage[dvipsnames]{xcolor}
\definecolor{deepfuchsia}{rgb}{0.76, 0.33, 0.76}
\definecolor{electricpurple}{rgb}{0.75, 0.0, 1.0}
\usepackage[colorlinks=true,linktoc=page,linkcolor=blue,citecolor=blue,urlcolor=electricpurple]{hyperref}

\usepackage{orcidlink}

\usepackage{titlesec} 

\usepackage[normalem]{ulem}
\usepackage{xcolor}

\titleformat{\section}
  {\raggedright\bfseries\large} 
  {\thesection}{1em}{}

\titleformat{\subsection}
  {\raggedright\bfseries\large} 
  {\thesubsection}{1em}{}

\titlespacing*{\section}
  {0pt} 
  {5pt} 
  {0pt} 

\titlespacing*{\subsection}
  {0pt} 
  {3pt} 
  {0pt} 

\newcommand{\beq}{\begin{equation}}
\newcommand{\eeq}{\end{equation}}
\newcommand{\bea}{\begin{eqnarray}}
\newcommand{\eea}{\end{eqnarray}}
\newcommand{\non}{\nonumber}

\newcommand{\eqn}[1] {Eq.~(\ref{#1})}
\newcommand{\fig}[1]{Fig.~\ref{#1}}
\newcommand{\mylabel}[1]{\label{#1}}

\usepackage{mathtools}
\DeclarePairedDelimiter\bra{\langle}{\rvert}
\DeclarePairedDelimiter\ket{\lvert}{\rangle}
\DeclarePairedDelimiterX\braket[2]{\langle}{\rangle}{#1 \delimsize\vert #2}

\renewcommand\thesection{\arabic{section}}
\renewcommand{\thesubsection}{\alph{subsection}}

\begin{document}

\title{Unconventional Superconductivity in the Chiral Topological Semimetal \texorpdfstring{Ag$_2$Pd$_3$S}{Ag2Pd3S}}

\author{{Roshan Kumar Kushwaha}\,\orcidlink{0009-0005-3457-3653}}\thanks{These authors contributed equally to this work}
\affiliation{Department of Physics, Indian Institute of Science Education and Research Bhopal, Bhopal, 462066, India}

\author{{Dibyendu Samanta}\,\orcidlink{0009-0004-3022-7633}}
\thanks{These authors contributed equally to this work}
\affiliation{Department of Physics, Indian Institute of Technology, Kanpur 208016, India}

\author{Sudarshan~Sharma\,\orcidlink{0000-0002-4710-9615}}
\affiliation{Department of Physics and Astronomy, McMaster University, Hamilton, Ontario L8S 4M1, Canada}

\author{Mathew~Pula\,\orcidlink{0000-0002-4567-5402}}
\affiliation{Department of Physics and Astronomy, McMaster University, Hamilton, Ontario L8S 4M1, Canada}

\author{{Shashank Srivastava}}
\affiliation{Department of Physics, Indian Institute of Science Education and Research Bhopal, Bhopal, 462066, India}

\author{{Poulami Manna}}
\affiliation{Department of Physics, Indian Institute of Science Education and Research Bhopal, Bhopal, 462066, India}

\author{{Arushi}}
\affiliation{Department of Physics, Indian Institute of Science Education and Research Bhopal, Bhopal, 462066, India}

\author{{Sajilesh K. P.}}
\affiliation{Physics Department, Technion-Israel Institute of Technology, Haifa 32000, Israel}

\author{{Suhani Sharma}}
\affiliation{Department of Physics, Indian Institute of Science Education and Research Bhopal, Bhopal, 462066, India}

\author{{Priya Mishra}}
\affiliation{Department of Physics, Indian Institute of Science Education and Research Bhopal, Bhopal, 462066, India}

\author{{Prabin Kumar Naik}}
\affiliation{Department of Physics, Indian Institute of Science Education and Research Bhopal, Bhopal, 462066, India}

\author{James Beare}
\affiliation{Department of Physics and Astronomy, McMaster University, Hamilton, Ontario L8S 4M1, Canada}

\author{Yipeng Cai}
\affiliation{TRIUMF, Vancouver, British Columbia V6T 2A3, Canada}
\affiliation{Quantum Matter Institute, The University of British Columbia, Vancouver, BC V6T 1Z4, Canada}

\author{Kenji M. Kojima}
\affiliation{TRIUMF, Vancouver, British Columbia V6T 2A3, Canada}

\author{{Amit Kanigel}}
\affiliation{Physics Department, Technion-Israel Institute of Technology, Haifa 32000, Israel}

\author{{Graeme M. Luke}}
\affiliation{Department of Physics and Astronomy, McMaster University, Hamilton, Ontario L8S 4M1, Canada}
\affiliation{TRIUMF, Vancouver, British Columbia V6T 2A3, Canada}

\author{{Sudeep Kumar Ghosh}\,\orcidlink{0000-0002-3646-0629}}
\email[]{skghosh@iitk.ac.in}
\affiliation{Department of Physics, Indian Institute of Technology, Kanpur 208016, India}

\author{{Ravi Prakash Singh}\,\orcidlink{0000-0003-2548-231X}}
\email[]{rpsingh@iiserb.ac.in} 
\affiliation{Department of Physics, Indian Institute of Science Education and Research Bhopal, Bhopal, 462066, India}

\begin{abstract}
Chiral crystals provide a unique setting where broken inversion symmetry, strong spin-orbit coupling, and electronic topology intertwine, yet superconductivity in intrinsically chiral materials remains rare. Here, we report unconventional superconductivity in the chiral topological semimetal Ag$_2$Pd$_3$S, an enantiomorphic analog of natural mineral coldwellite, crystallizing in the right-handed space group $P4_132$. Bulk superconductivity with a transition temperature $T_C = 1.1(2)$ K is confirmed by electrical resistivity, magnetization, and specific-heat measurements. Muon spin rotation and relaxation ($\mu$SR) experiments reveal a fully gapped superconducting state that spontaneously time-reversal symmetry (TRS) breaking establishing Ag$_2$Pd$_3$S as the first chiral topological semimetal superconductor exhibiting intrinsic TRS breaking. First-principles calculations uncover multiple multifold band crossings near the Fermi level, hosting Kramers-Weyl, double spin-1, and spin-3/2 quasiparticles with large topological charges. These unconventional fermions generate symmetry-protected topological surface states and underscore the nontrivial topology of the normal state. Symmetry analysis based on the Ginzburg-Landau theory suggests a loop-supercurrent-ordered superconducting state, yielding a full gap alongside spontaneous TRS breaking. The coexistence of TRS-breaking superconductivity and chiral multifold fermions identifies Ag$_2$Pd$_3$S as a platform for realizing intrinsic superconducting diode effects and chirality-induced spin selectivity, offering a transformative pathway toward dissipationless topological quantum technologies.
\end{abstract}

\maketitle

\section{Introduction}
Chirality, the geometric property that distinguishes an object from its mirror image, is a foundational concept in the natural sciences, from molecular biology to the physics of quantum materials~\cite{chiral}. In crystalline solids, structural chirality emerges when inversion, mirror, and roto-inversion symmetries are absent, resulting in handed atomic arrangements belonging to the 65 Sohncke space groups~\cite{sohncke1879}. These crystals exhibit a wide variety of symmetry-driven responses, including piezoelectricity, natural optical activity, circular dichroism, and nonreciprocal charge transport~\cite{multunas2023circular,ideue2017bulk}. More recently, the interplay between structural chirality, spin-orbit coupling (SOC), and electronic topology has attracted considerable attention, as it can stabilize unconventional quasiparticles such as Kramers-Weyl and multifold fermions carrying large topological charges~\cite{bradlyn2016beyond,menichetti2025chirality}. Unlike their counterparts in high-energy physics, these quasiparticles are protected by the crystal's chiral symmetries, leading to striking phenomena including extended helicoid Fermi arcs, strong Berry curvature effects, and quantized circular photogalvanic responses~\cite{multunas2023circular}. Prototypical examples include the B20-type compounds CoSi and RhSi, where multifold fermions and their associated topological surface states have been experimentally confirmed~\cite{B20}. The combination of chirality, topology, and SOC thus provides a fertile setting for the realization of novel quantum phases of matter.

\begin{figure*} [!ht] 
\includegraphics[width=\textwidth,origin=b]{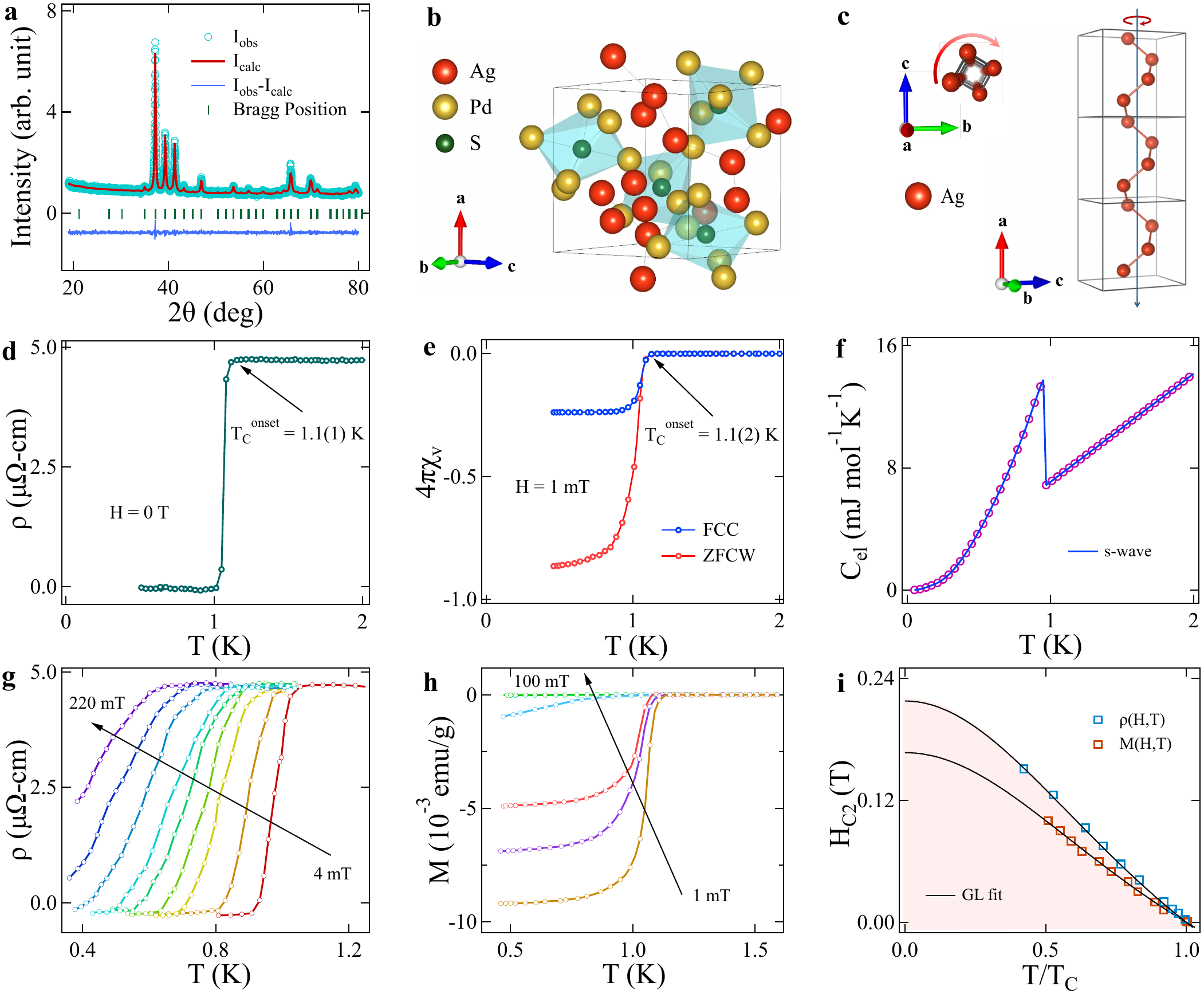}
\caption{\label{crystal} \textbf{Crystal structure and bulk superconductivity}: a) Rietveld-refined powder XRD pattern. b) Cubic crystal structure of Ag$_{2}$Pd$_{3}$S. c) Top view (left) and side view (right) shows of right-handed Ag-helix. The superconducting transition temperature of Ag$_{2}$Pd$_{3}$S is confirmed by temperature-dependent d) electrical resistivity, e) magnetization, and f) electronic specific heat, with the solid blue line showing the s-wave model fit. The upper critical field is evaluated using temperature-dependent g) resistivity, and h) magnetization. i) GL fit describing the variation of upper critical field with $T/T_C$.}
\end{figure*}

Because chiral crystals are inherently noncentrosymmetric (NC), they provide a specialized platform for exploring unconventional superconductivity driven by the absence of inversion symmetry~\cite{Smidman2017,bauer2012non}. In such systems, antisymmetric spin-orbit coupling (ASOC) lifts spin degeneracy and allows the admixture of spin-singlet and spin-triplet pairing channels~\cite{Smidman2017,bauer2012non,ASOC,CePt3Si}, leading to unusual properties such as nonreciprocal transport and anomalous optical responses~\cite{Nagaosa2024}. A particularly striking phenomenon observed in a small subset of NC superconductors is time-reversal symmetry breaking in the superconducting state~\cite{Ghosh2020review,barker2015unconventional,singh2020time,arushi2021time,Shang2022spin,singh2014detection,singh2017time,singh2018time,Hillier2009,Shang2020time,Shang2018,shang2020simultaneous,sharma2023evidence,Sajilesh2024hfrhge}. However, despite significant theoretical interest, superconductivity in chiral crystals remains comparatively rare. Most known examples, B20 type material, such as AuBe, RhGe and BiPdSe, a B20 materials, exhibit conventional $s$-wave behavior~\cite{B20}, and only a few compounds show signatures of mixed-parity pairing linked to chiral fermions~\cite{gao2022topological}. Although metallic systems with nontrivial band topology are proposed as promising platforms for topological superconductivity~\cite{srivastava2026hourglass,Pavan_2025,Pavan_2025_Zr,kataria2026,badger2022dirac,sudeepZrOsSi,Ghosh2020b,Yadav2024,Ghosh2022Dirac}, the microscopic mechanisms through which spin-momentum locking and Fermi surface topology of chiral crystals influence pairing symmetry and TRS-breaking remain open questions. This motivates the search for new materials where topology and symmetry intertwine at both the electronic and lattice levels, raising the fundamental question: can a chiral topological material host unconventional superconductivity with broken time-reversal symmetry and potentially realize topological superconductivity?

Here we address this question by investigating a chiral compound, Ag$_2$Pd$_3$S. It crystallizes in the right-handed chiral space group $P4_132$ (No. 213), featuring intertwined Pd-S and Ag-S helical networks that support multifold band crossings near the Fermi level. The right-handed compound is the enantiomeric form of the natural mineral coldwellite, discovered in the Coldwell Complex of Ontario, Canada. Low-temperature resistivity, magnetization, and specific-heat measurements confirm bulk superconductivity with a transition temperature $T_C = 1.1(2)$ K, consistent with earlier reports~\cite{mcdonald2015coldwellite,Ag2Pd3SAPS,SchaakJACS,APSold}. Transverse-field muon spin rotation/relaxation ($\mu$SR) reveals a fully gapped superconducting state, whereas zero-field $\mu$SR measurements confirm Ag$_2$Pd$_3$S as the first chiral topological semimetal superconductor with intrinsic time-reversal symmetry (TRS) breaking. Complementary first-principles calculations uncover a complex topological landscape featuring multiple multifold fermionic excitations near the Fermi level, including Kramers-Weyl, double spin-1, and spin-3/2 quasiparticles with distinct topological charges. The coexistence of these exotic large topological charge quasiparticles with a TRS-breaking superconducting state identifies Ag$_2$Pd$_3$S as a rare system where chirality, topology and superconductivity converge. The superconducting phase is consistent with a loop-supercurrent order state, suggesting that Ag$_2$Pd$_3$S provides a promising platform for exploring intrinsic chiral topological superconductivity and related phenomena such as superconducting diode effects~\cite{ma2025superconducting, Nadeem2023} and chirality-induced spin selectivity~\cite{bloom2025using}.

\section{Results}

\noindent\textbf{Chiral Crystal Structure of Ag$_{2}$Pd$_{3}$S:}

The Rietveld-refined powder X-ray diffraction pattern identifies the sample in a pure phase (see Fig.~\ref{crystal}(a)). It crystallizes in a cubic $\beta$-Mn-type chiral structure with the space group $P4_132$, as shown in Fig.~\ref{crystal}(b). The structure is quite analogous to Li$_2$M$_3$B (M = Pd, Pt) (space group: $P4_332$); however, these two structures exhibit distinct types of crystallographic chirality: Ag$_{2}$Pd$_{3}$S and Li$_2$M$_3$B have right- and left-handed chiral crystal structures, respectively. The $\beta$-Mn structure provides an accurate representation of this crystal structure, where Ag, Pd, and S atoms occupy the 8c, 12d, and 4a sites. The shaded polyhedra in Fig.~\ref{crystal}(b) correspond to Pd$_6$S octahedra, with a sulfur atom in the center. The interstitial voids within the Ag network are occupied by corner-sharing Pd$_6$S octahedra~\cite{mcdonald2015coldwellite,Ag2Pd3SAPS,SchaakJACS,APSold}. The right-handed Ag-helices are present along the [100], [010], and [001] directions in Ag$_{2}$Pd$_{3}$S. Fig.~\ref{crystal}(c) shows the right-handed Ag-helix along the [100] direction.\\

\noindent\textbf{Bulk Superconductivity in Ag$_{2}$Pd$_{3}$S:}

\textit{\textbf{Resistivity:}}
Low-temperature resistivity measurements on Ag$_{2}$Pd$_{3}$S show a sudden drop in resistivity at the onset of 1.1(1)~K, confirming the superconducting transition, as shown in Fig.~\ref{crystal}(d). Fig.~\ref{crystal}(g) shows the field-dependent resistivity data ($\rho(T, H)$), where the transition temperature is suppressed as the magnetic field increases. The upper critical field was calculated using the Ginzburg-Landau (GL) equation as shown in Fig.~\ref{crystal}(i) (details in the supplementary information).\\

\textit{\textbf{Magnetization:}}
Magnetization measurements were carried out in two modes: zero field cooled warming (ZFCW) and field cooled cooling (FCC), in an applied field of 1~mT, revealing a superconducting transition temperature of $T_{C}$ = 1.1(2)~K as shown in Fig.~\ref{crystal}(e). Field and temperature-dependent magnetization measurements (see Fig.~\ref{crystal}(h) and Fig.~S2, supplementary information) were performed to calculate the lower critical fields ($H_{C1}$) and the upper critical fields ($H_{C2}$). The plot of $H_{C2}$ versus temperature is shown in Fig.~\ref{crystal}(i). The detailed analysis and evaluation of other superconducting parameters are discussed in the supplementary information.\\

\begin{figure*}[t!] 
\includegraphics[width=\textwidth,origin=b]{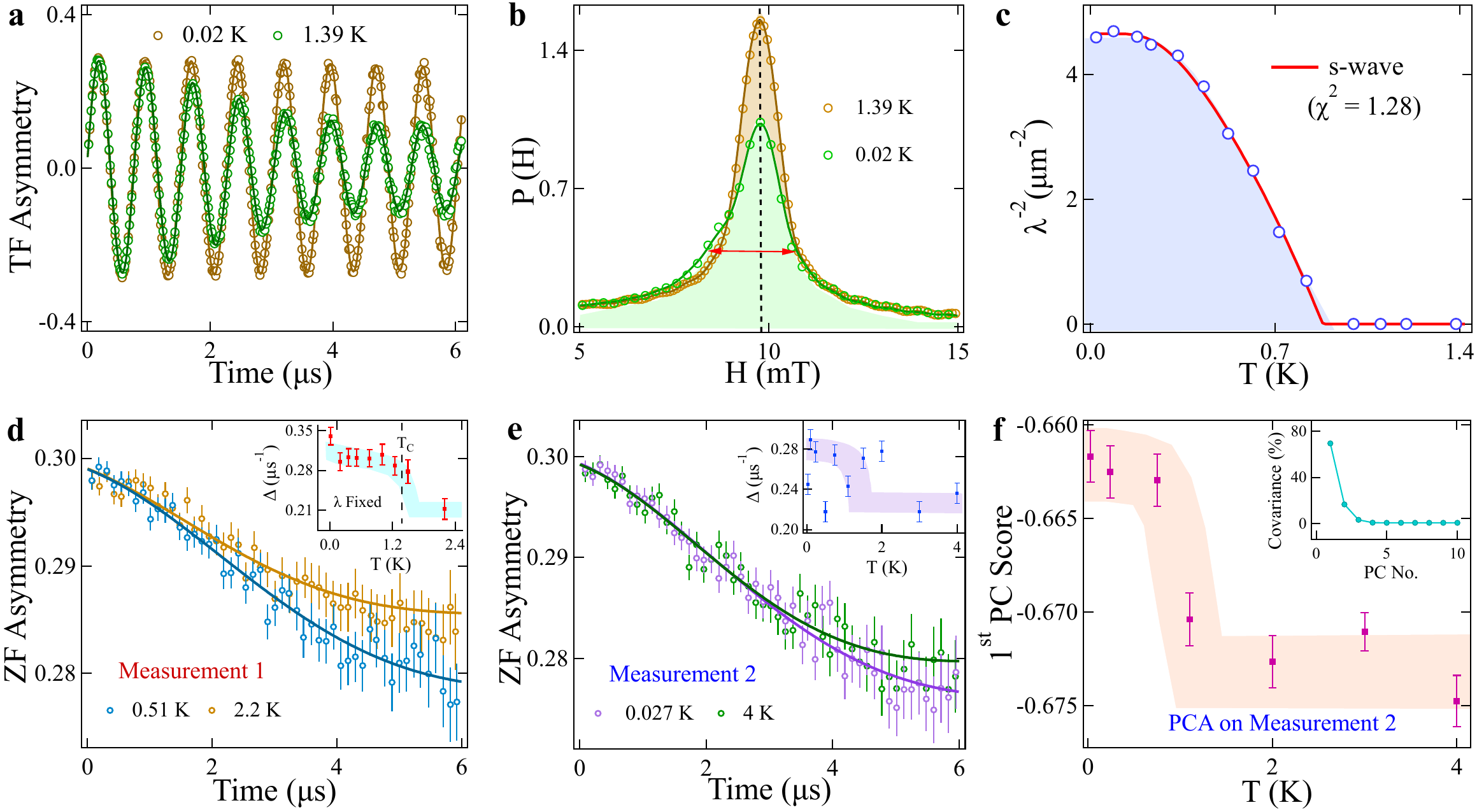}
\caption{\label{muon} \textbf{Microscopic investigation via $\mu$SR and Principal Component Analysis (PCA)}: a) TF-asymmetry spectra recorded at 0.02~K and 1.39~K. b) Broadening in the field distribution below $T_C$ and sharp distribution above $T_C$. c) Temperature dependence of $\lambda^{-2}$, where the solid red line represents the fit to the s-wave model. ZF-asymmetry spectra recorded above and below $T_C$ for d) measurement 1, and e) measurement 2, where the insets show variation of nuclear relaxation rate ($\Delta$) with temperature for respective measurements. f) PCA on measurement 2 shows a subtle increment in the 1$^{st}$ PC score with temperature. The inset shows the covariance captured by different PCs.}
\end{figure*}

\textit{\textbf{Specific heat:}}
The sharp jump in the specific heat ($C_P$) data at 1.1(2)~K reveals the superconducting transition temperature (Fig. S3, supplementary information). The normal-state region of the total specific heat is fitted to the Debye-Sommerfeld model, as described in the supplementary information. Furthermore, the inverse McMillan's model~\cite{McM} estimates weak electron-phonon coupling strength in Ag$_{2}$Pd$_{3}$S. Fig.~\ref{crystal}(f) represents the temperature-dependent electronic specific heat ($C_{el}$) at zero field. The pairing symmetry of the superconducting gap can be analyzed using the temperature dependence of the electronic specific heat ($C_{el}$). The s-wave model provides the best fit, as shown by the solid blue curve in Fig.~\ref{crystal}(f). The normalized electronic specific heat jump $\Delta C_{el}/\gamma T_{C}$ = 1.03, which is less than the value for the BCS superconductors (1.43). Details of the analysis of gap symmetry are provided in the supplementary information.\\

\noindent\textbf{Microscopic Investigation through $\mu$SR:}

The results of transverse-field (TF) and zero-field (ZF)-$\mu$SR experiments are discussed below:

\textit{\textbf{Analysis of gap symmetry:}} In particular, TF-$\mu$SR measurements allow us to extract the magnetic penetration depth ($\lambda(T)$), which is directly related to the superfluid density, and hence the gap symmetry. By analyzing the temperature dependence of the muon depolarization rate, we can assess the symmetry of the superconducting  gap symmetry. It is crucial to identify whether the superconducting pairing in Ag$_2$Pd$_3$S is conventional or influenced by the underlying chiral crystal symmetry and the multifold fermionic band topology. The sample was field-cooled under a magnetic field, $H=10$~mT ($H_{C1}<H<<H_{C2}$). The asymmetry spectra recorded below $T_{C}$ (0.02~K) and above $T_{C}$ (1.39~K) are presented in Fig.~\ref{muon}(a). The difference in the depolarization rate between the $T < T_{C}$ spectra and those above $T_{C}$ can be explained by an inhomogeneous field distribution arising from flux line lattice (FLL) formation. The weak Gaussian damping for the $T > T_{C}$ spectra corresponds to randomly oriented nuclear dipolar fields. The distributions of the field in the vortex and normal states are shown in Fig.~\ref{muon}(b), where a peak in the internal field is present at the applied field value in the normal state. A small hump is present in the vortex state (0.02 K) below the applied field peak due to the formation of FLL. The time-domain asymmetry is determined by two oscillatory functions multiplied by Gaussian and exponential relaxation functions, corresponding to the sample and background contribution from the silver sample holder, respectively, as shown in Eq.~\ref{ATF}.
\begin{equation}
\begin{split}
A_{TF}(t) = A_{i} [f_{s} \mathrm{exp} \left (\frac{- \sigma^{2} t^{2}}{2}\right) \mathrm{cos}(\omega t+ \phi)\\ + (1-f_{s}) \mathrm{exp}(- \psi t) \mathrm{cos} (\omega_{bg} t +\phi)],
\end{split}
\label{ATF}
\end{equation}
where $A_{i}$ is the initial asymmetry and $f_{s}$ is the signal corresponding to the sample contribution. $\sigma$ and $\psi$ are the relaxation rates of the sample and the background. $\phi$ is the initial phase, $\omega$ and $\omega_{bg}$ are the precession frequencies corresponding to the muon implanted in the sample and the sample holder, respectively. The total depolarization, $\sigma$, includes the contribution of FLL in the superconducting state ($\sigma_{sc}$) and a tiny contribution from randomly oriented nuclear dipole moments ($\sigma_{n}$), which is represented as
\begin{equation}
{\sigma}^2 = {\sigma_{sc}}^2 +{\sigma_{n}}^2.
\label{sigsc}
\end{equation}
We present $\sigma$ as a function of temperature for Ag$_{2}$Pd$_{3}$S obtained from the fitting (see the supplementary information). The formation of FLL in the superconducting state due to an inhomogeneous field distribution inside the sample corresponds to a smooth enhancement in $\sigma$ below $T_{C}$. The background nuclear contribution to the depolarization rate $\sigma_{n}$ = 0.082(7)~$\mu s^{-1}$ was found to be temperature independent above $T_{C}$ and using Eq.~\ref{sigsc} the superconducting contribution to the relaxation rate $\sigma_{sc}$ was extracted and found to be 0.4924(3)~$\mu s^{-1}$ at T = 0~K. $\sigma_{sc}$ depends on the second moment of the internal magnetic field related to the mean square inhomogeneity in the field ($<(\Delta B)^{2}>$) as ${\sigma_{sc}}^{2} = {\gamma_{\mu}}^{2} <(\Delta B)^{2}>$, where $\gamma_{\mu} = 2\pi \times$135.5 MHz/T is the gyromagnetic ratio of the muon. For a perfect triangular/square vortex lattice, the penetration depth can be calculated from the relaxation rate using the formula~\cite{EHB}
\begin{equation}
\label{lambdasc}
\sigma_{sc}(T) = \frac{4.854 \times10^4 (1-b)\left[1+1.21 \left(1-\sqrt{b}\right)\right]}{{\lambda}^{2}(T)},
\end{equation}
where b is the reduced magnetic field ($b = \langle H \rangle_{appl}/H_{c2}$) and $\sigma_{sc}$ is in units of $\mu s^{-1}$ and $\lambda$ is in nm. Here, the magnetization derived value $H_{c2}$ is used to calculate $\lambda(T)$ from $\sigma_{sc}(T)$. The magnetic field applied in the TF geometry was $\langle B \rangle_{appl}$ = 10~mT. Eq.~\ref{lambdasc} is applicable for type-II superconductors with $b$ $\leq$ 0.25 and $k_{GL}$ $\geq$ 5~\cite{ZrIrSi, CaPd2Ge2}. The temperature variation of $\lambda^{-2}$ is shown in Fig.~\ref{muon}(c) and is described by Eq.~\ref{lambdasc}, which yields $\sigma_{sc}$(0)= 0.4924(5)~$\mu$$s^{-1}$ and the corresponding magnetic penetration depth at 0~K was estimated as $\lambda$(0) = 376(5)~nm, which is close to the GL penetration depth ($\lambda_{GL}(0)=369(6)$~nm, see supplementary information) estimated from low-temperature magnetization measurement. The s-wave model nicely fits the $\lambda^{-2}(T)$ data as shown in Fig.~\ref{muon}(c) with a solid red curve. A detailed analysis of gap symmetry using the s-wave model is described in the supplementary information. Both the specific-heat and magnetic penetration-depth results from TF-$\mu$SR consistently indicate the presence of a fully-gapped isotropic superconducting gap in Ag$_{2}$Pd$_{3}$S.

\textit{\textbf{Investigation of time reversal symmetry:}}
ZF-$\mu$SR measurements were performed to detect spontaneous magnetic fields, indicating the presence of TRS-breaking in the superconducting state. Relaxation spectra were collected in the absence of any external magnetic field, below $T_C$ (0.02~K) and above $T_C$ (4.0~K), in two separate measurements: measurement 1 (Fig.~\ref{muon}(d)) and measurement 2 (Fig.~\ref{muon}(e)). There were no oscillatory components in the spectra, which indicates the absence of any magnetic ordering. Due to randomly oriented nuclear moments, relaxation spectra are generally modeled by the Gaussian Kubo-Toyabe (KT) function \cite{KT}, where there are no static electronic moments.
\begin{equation}
\label{KT}
G_{KT}(t)= \frac{1}{3}+\frac{2}{3} (1- {\Delta}^{2}t^{2}) e^{{-\Delta}^{2}t^{2}/2},
\end{equation}
where $\Delta_{ZF}$ is the static Gaussian relaxation rate due to randomly oriented nuclear dipole moments. The following function describes the time evolution of ZF asymmetry spectra:
\begin{equation}
\label{AZF}
A(t) = A_{i} G_{KT}(t) e^{-\Lambda t}+ A_{BG},
\end{equation}
where $\Lambda$ is the electronic relaxation rate associated with the electronic moments, and $A_{BG}$ is the background contribution to the asymmetry spectra of the muons stopped in the sample holder.

\begin{figure*} [!htb] 
\includegraphics[width=\textwidth]{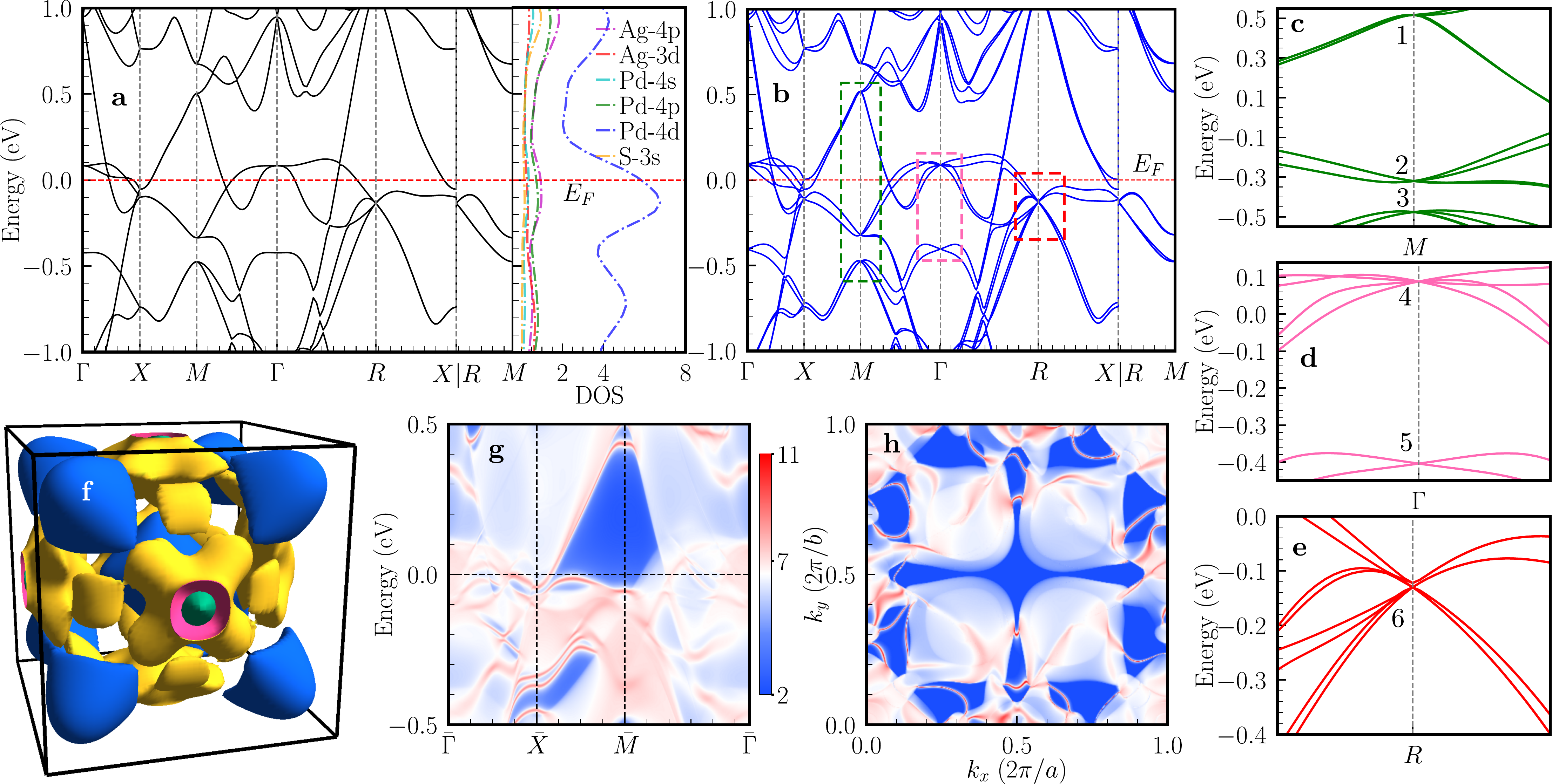}
\caption{\label{fig:band_topology} \textbf{Electronic band structure, multifold fermions, and surface states of ${\rm Ag}_2{\rm Pd}_3{\rm S}$:} a) Electronic band structure and corresponding orbital-resolved density of states of ${\rm Ag}_2{\rm Pd}_3{\rm S}$ calculated without spin-orbit coupling (SOC). b) Electronic band structure of ${\rm Ag}_2{\rm Pd}_3{\rm S}$ with SOC included.  (c-e) Four-fold double Kramers-Weyl fermions at $M$ point, four-fold Rarita-Schwinger-Weyl fermions, and two-fold Kramers-Weyl at point $\Gamma$, six-fold double spin-$1$ fermion at $R$ point are depicted. f) Fermi surface of ${\rm Ag}_2{\rm Pd}_3{\rm S}$ including SOC, showing multiple sheets across the Brillouin zone. g) Surface state spectrum of ${\rm Ag}_2{\rm Pd}_3{\rm S}$ along high-symmetry paths of the projected (001) two-dimensional surface Brillouin zone. h) Constant-energy slice of the surface spectrum at $\sim -0.02$ eV, revealing topological surface Fermi arcs.}
\end{figure*}

In measurement 1, there is a clear difference in the asymmetry spectra above and below $T_C$, suggesting the possibility of TRS-breaking. The fitting of ZF-asymmetry data using Eq.~\ref{AZF} reveals an enhancement in the temperature-dependent nuclear relaxation rate ($\Delta$) just below $T_C$ (see the inset of Fig.~\ref{muon}(d)). To verify the reproducibility of this spontaneous internal magnetic field, measurement 2 conducted. Here, the ZF-asymmetry spectra again show variation with temperature, yielding relaxation rates that evolve systematically across $T_C$ as shown in the inset of Fig.~\ref{muon}(e). The observations of both measurements point to the possible presence of TRS-breaking in Ag$_{2}$Pd$_{3}$S. 

To gain further insight into the ZF-$\mu$SR measurements of ${\rm Ag}_2{\rm Pd}_3{\rm S}$, we employed Principal Component Analysis (PCA)~\cite{geron2022hands}, an unsupervised machine-learning method used to extract dominant trends in high-dimensional data, as illustrated in \fig{muon}(f). In this approach, the measured asymmetry spectra are represented as a set of orthogonal basis vectors, or principal components (PCs), that capture the leading sources of variance in the dataset. The associated PC projections (scores) trace systematic temperature-dependent variations in the $\mu$SR asymmetry signal. PCA has previously proven effective in identifying TRS-breaking superconducting transitions in materials such as LaNiGa$_2$ and LaNi$_{1-x}$Cu$_x$C$_2$, as well as magnetic ordering in BaFe$_2$Se$_2$O~\cite{tula2021machine}. While PCA efficiently highlights subtle spectral changes associated with such transitions, it does not inherently distinguish which specific relaxation channels are responsible for the observed variations. 

For ${\rm Ag}_2{\rm Pd}_3{\rm S}$, the second ZF-$\mu$SR measurement reveals a slight change in the asymmetry spectra in $T_C$. To further evaluate the data, we performed a joint PCA with other known TRS-breaking superconductors~\cite{tula2021machine}. The scree plot in the inset of \fig{muon}(f) shows that the first two PCs capture the dominant variance, with the first PC contributing the most significantly. The corresponding PC score uncertainties were estimated from the experimental errors in the asymmetry functions~\cite{tula2021machine}. The temperature dependence of the PC scores, shown in \fig{muon}(f), reveals a pronounced change in the first PC for $T \lesssim T_C$, indicating the onset of spontaneous time-reversal symmetry breaking in ${\rm Ag}_2{\rm Pd}_3{\rm S}$.\\

\noindent\textbf{Topology of electronic band structure:} 

The electronic band structure of ${\rm Ag}_2{\rm Pd}_3{\rm S}$ was calculated using the QUANTUM ESPRESSO ab initio simulation package, based on density functional theory (DFT) \cite{Giannozzi_2009,Giannozzi_2017,Giannozzi_2020}, with the Perdew-Burke-Ernzerhof (PBE) functional employed to approximate the exchange-correlation potential \cite{Perdew_1996}. The band structure computed without spin-orbit coupling (SOC), shown in Fig.~\ref{fig:band_topology}(a), exhibits several dispersive bands crossing the Fermi level, resulting in both electron and hole pockets and reflecting the multiband nature of ${\rm Ag}_2{\rm Pd}_3{\rm S}$. The low-energy bands near the Fermi level are primarily composed of Pd-4d orbitals, as confirmed by the orbitally resolved projected density of states, also shown in Fig.~\ref{fig:band_topology}(a), suggesting a strong influence of antisymmetric SOC. Upon inclusion of SOC, significant band splittings are observed near the Fermi level, as shown in Fig.~\ref{fig:band_topology}(b). Remarkably, several symmetry-protected band crossings survive in the presence of SOC as a result of nonsymmorphic crystal symmetries, exhibiting unconventional degeneracies beyond the standard twofold (Weyl) and fourfold (Dirac) cases. The two-fold Kramers-Weyl node on the label $(5)$, carrying monopole charge $\pm 1$, three four-fold double Kramers-Weyl nodes on labels $(1)$, $(2)$, and $(3)$ with monopole charge $\pm 2$, a six-fold double spin-$1$ fermion associated with monopole charge $\pm 4$ on label $(6)$, and four-fold spin-$3/2$ fermion on label $(4)$ which also possess monopole charge $\pm 4$. These exotic quasiparticles are distributed across high-symmetry points of the Brillouin zone, as indicated in Fig.~\ref{fig:band_topology}, and underline the rich topological character of the electronic structure. Furthermore, the calculated Fermi surface of ${\rm Ag}_2{\rm Pd}_3{\rm S}$, shown in Fig.~\ref{fig:band_topology}(c), displays a rich structure characteristic of its multiband topology. It consists primarily of ellipsoid-like electron and hole pockets, together with a large $\Gamma$-centered surface that is topologically equivalent to a hollow sphere perforated along the cubic axes, resembling a ``wiffle ball''. In addition, distinct lens-shaped pockets are observed along the $\langle 111 \rangle$ directions. Together, these features produce a Fermi surface in which the pockets enclose all time-reversal invariant momenta (TRIMs), providing compelling evidence for the nontrivial topological character of this material.

${\rm Ag}_2{\rm Pd}_3{\rm S}$ has a cubic structure and belongs to the chiral space group ${\rm P}4_132$ (No 213). Its generators contain the following rotation operations (screw): $\{C_{2x}|\frac{1}{2}\frac{1}{2}0\}$, $\{C_{2y}|0\frac{1}{2}\frac{1}{2}\}$, $\{C_{3,111}^{-}|000\}$, and $\{C_{2,1\bar{1}0}|\frac{3}{4}\frac{3}{4}\frac{3}{4}\}$. At the $\Gamma$-point, the full cubic symmetry is preserved, and the little group admits four-dimensional irreducible representations once SOC is included~\cite{bradlyn2016beyond}. As a result, bands are enforced to form four-fold degeneracies, corresponding to unconventional fermions such as spin-$3/2$ (Rarita-Schwinger-Weyl) fermions with monopole charge $\pm 4$. In contrast, at the zone-corner $R=(\pi, \pi, \pi)$, nonsymmorphic screw-rotation symmetries enforce three-dimensional irreducible representations~\cite{bradlyn2016beyond}. Since the $R$-point is invariant under time-reversal symmetry, each of these three states is doubled by the Kramers theorem, yielding six mutually orthogonal states. This mechanism guaranties the presence of six-fold degeneracies in $R$, which correspond to double spin-$1$ fermions associated with monopole charge $\pm 4$. Thus, the four-fold fermions at $\Gamma$ and six-fold fermions at $R$ are not accidental band crossings, but rather symmetry-enforced features of the cubic nonsymmorphic space group.

Finally, we investigate the electronic surface structure of ${\rm Ag}_2{\rm Pd}_3{\rm S}$ by projecting the bulk band structure onto the $(001)$ surface Brillouin zone as shown in Fig.~\ref{fig:band_topology}(g). The resulting surface spectrum reveals several topologically nontrivial surface states crossing the Fermi level. In particular, projection of the sixfold degenerate node at the $R$ point, along with several fourfold-degenerate nodes at the $M$ point, onto the $(001)$ surface gives rise to surface Fermi arcs that resemble drumhead-like states, as shown in Fig.~\ref{fig:band_topology}(h). Additionally, projection of the fourfold-degenerate node at the $\Gamma$ point onto the same surface (001) produces a distinct surface Fermi arc pattern, illustrated in Fig.~\ref{fig:band_topology}(i). Notably, a prominent and extended Fermi arc connects the surface high-symmetry points $\bar{\Gamma}$ and $\bar{M}$, providing a clear signature of the nontrivial topological character of these multifold fermions.\\

\section{Discussion}
TRS-breaking in superconductors can originate from different mechanisms depending on whether the superconducting order parameter develops a complex structure in momentum space or in real space. In the former case, TRS-breaking typically arises from multicomponent order parameters belonging to higher-dimensional irreducible representations of the crystal symmetry group, which often leads to nodal quasiparticle excitations~\cite{Annett1990,Sigrist1991}. In contrast, real-space mechanisms can generate TRS-breaking without requiring nodes in the superconducting gap~\cite{Ghosh2020review}. To understand the nature of the superconducting ground state in ${\rm Ag}_2{\rm Pd}_3{\rm S}$, we therefore perform a symmetry analysis within the framework of Ginzburg-Landau (GL) theory~\cite{Annett1990,Sigrist1991,Ghosh2020review}. ${\rm Ag}_2{\rm Pd}_3{\rm S}$ is a chiral multifold semimetal crystallizing in the noncentrosymmetric cubic space group $P4_132$ (No. 213), corresponding to the point group $\mathcal{O}$ (432). Within GL theory, the normal-state symmetry group can be written as $\mathcal{G} = G \otimes U(1) \otimes \mathcal{T}$, where $G$ represents the combined spatial and spin rotational symmetries, $U(1)$ accounts for gauge invariance, and $\mathcal{T}$ denotes time-reversal symmetry. The cubic point group $\mathcal{O}$ contains five irreducible representations: two one-dimensional ($A_1$, $A_2$), one two-dimensional ($E$) and two three-dimensional ($T_1$, $T_2$). Among these, only the multidimensional representations can support complex order parameters that spontaneously break time-reversal symmetry.\\

\begin{figure}[!t]
\includegraphics[width=1.0\columnwidth,origin=b]{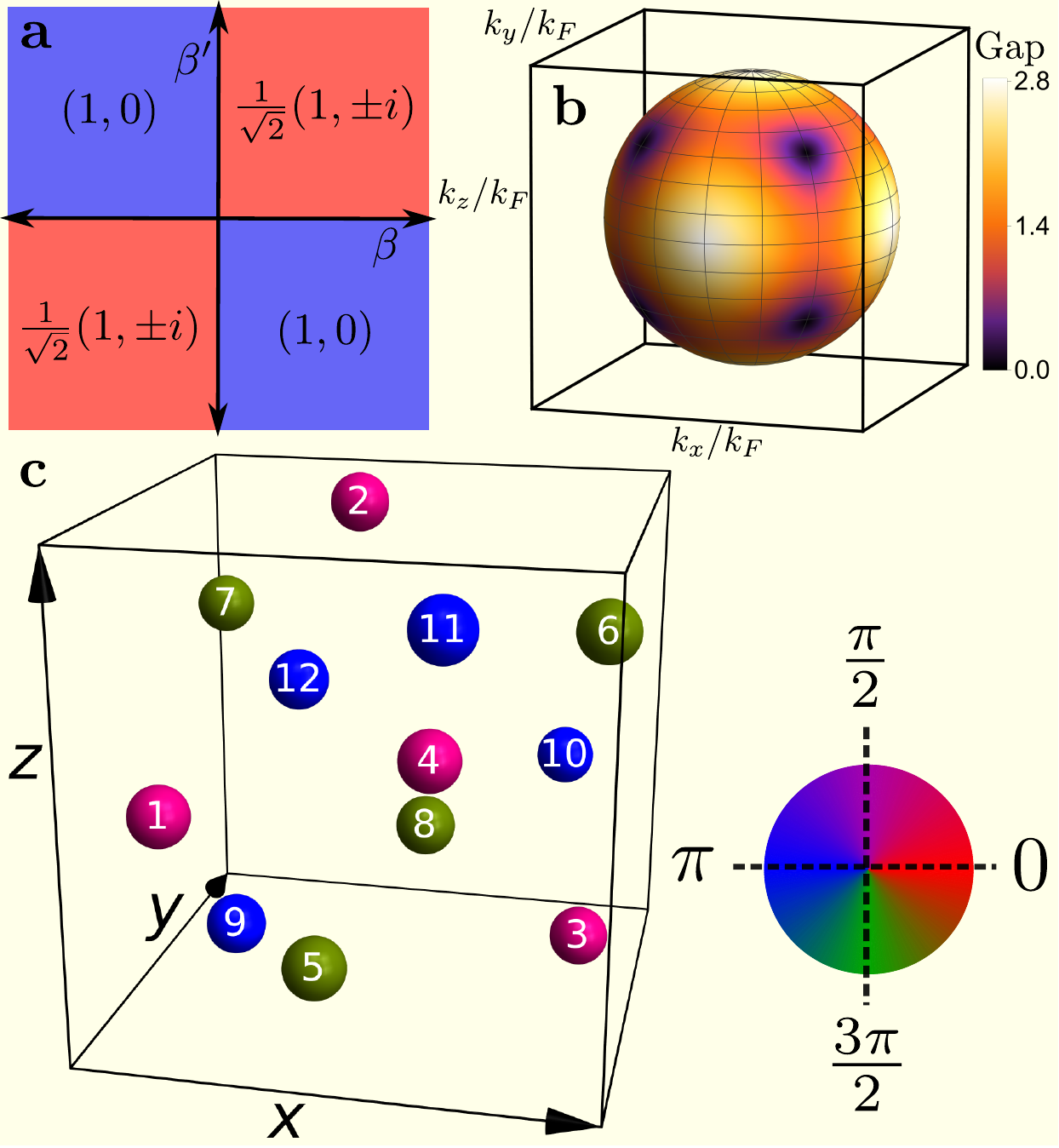}
\caption{\textbf{Nature of the superconducting ground state in ${\rm Ag}_2{\rm Pd}_3{\rm S}$:} a) GL phase diagram showing the symmetry-allowed two-component order parameters $(\eta_1,\eta_2)$ corresponding to the two-dimensional irreducible representation (irrep) $E$ of the cubic point group $O$. The TRS-breaking states $(\eta_1,\eta_2)=\tfrac{1}{\sqrt{2}}(1,\pm i)$ emerge when the ratio of GL parameters satisfies $\beta'/\beta>0$. b) Polar plots of the lowest quasiparticle excitation gap illustrating mixed singlet-triplet TRS-breaking superconducting states on a spherical Fermi surface. c) Visualization of the superconducting order parameter distribution in the loop-supercurrent ground state of ${\rm Ag}_2{\rm Pd}_3{\rm S}$. Each sphere denotes one of the 12 symmetrically distinct Pd sites within the unit cell, and the color wheel represents the complex phase of the loop supercurrent order parameter.}
\label{fig:LSC}
\end{figure}

We first consider superconducting instabilities allowed by the two-dimensional irrep $E$, for which the order parameter with complex components is written as $(\eta_1,\eta_2)$. Symmetry restricts the quartic part of the GL free energy to~\cite{Annett1990,Sigrist1991,Ghosh2020review}

\beq
f_4^{(1)}  =  \beta \left(|\eta_1|^2 + |\eta_2|^2\right)^2 + \beta^\prime \left(\eta^*_1 \eta_2 - \eta_1 \eta^*_2\right)^2,
\label{eqn:GL_fenergy}
\eeq

where $\beta$ and $\beta'$ are material-dependent GL coefficients. Minimization of the free energy in \eqn{eqn:GL_fenergy} yields two symmetry-allowed superconducting states: $(\eta_1, \eta_2) = (1, 0)$ and $\tfrac{1}{\sqrt{2}}(1, \pm i)$. The latter corresponds to a TRS-breaking superconducting state, which is energetically favored for an extended region in the parameter space as shown in \fig{fig:LSC}(a). Assuming strong SOC effects, relevant for ${\rm Ag}_2{\rm Pd}_3{\rm S}$, the TRS-breaking order parameters for the $E$ irrep can be expressed in the singlet channel $\Delta^{(1)}_s(\mathbf{k}) = A_1\left[(2k_z^2 - k_x^2 - k_y^2) + i\sqrt{3}(k_x^2 - k_y^2)\right]$ and in the triplet channel $\mathbf{d}^{(1)}(\mathbf{k}) = A_2\left[(i\sqrt{3} - 1)k_x, -(i\sqrt{3} + 1)k_y, 2 k_z\right]$, where $A_1$ and $A_2$ are material-dependent constants.\\

We next examine the three-dimensional irreps $T_1$ and $T_2$, for which the superconducting order parameter has three complex components $(\eta_1,\eta_2,\eta_3)$. The symmetry-allowed quartic GL free energy takes the form

\bea
f^{(2)}_4 &=& \beta \left(|\eta_1|^2 + |\eta_2|^2 + |\eta_3|^2\right)^2 + \beta' |\eta_1^2 + \eta_2^2 + \eta_3^2|^2 \non \\
&+& \beta'' \left(|\eta_1|^2 |\eta_2|^2 + |\eta_2|^2 |\eta_3|^2 + |\eta_3|^2 |\eta_1|^2\right),
\eea

where $\beta$, $\beta'$, and $\beta''$ are material-dependent GL coefficients. Minimizing this free energy with respect to $\eta_1$, $\eta_2$, and $\eta_3$ yields several competing superconducting states, some of which break TRS. One such TRS-breaking configuration corresponds to $\tfrac{1}{\sqrt{2}}(1, \pm i, 0)$. The corresponding symmetry-allowed superconducting order parameters for the $T_1$ irrep are given by the singlet component $\Delta^{(2)}_s(\mathbf{k}) = A_3 [k_y k_z (k_y^2 - k_z^2) + i k_z k_x (k_z^2 - k_x^2)]$ and the triplet component $\mathbf{d}^{(2)}(\mathbf{k}) = A_4 [-i k_z, k_z, (i k_x - k_y)]$. Similarly, for $T_2$ irrep, the TRS-breaking singlet and triplet components take the forms $\Delta^{(3)}_s(\mathbf{k}) = A_5 [k_x k_y + i k_x k_z]$ and $\mathbf{d}^{(3)}(\mathbf{k}) = A_6 [i k_z, k_z, (i k_x + k_y)]$, respectively. Here, $A_i$ are material-dependent constants independent of $\mathbf{k}$. Both $\mathbf{d}^{(1)}(\mathbf{k})$ and $\mathbf{d}^{(2)}(\mathbf{k})$ represent nonunitary triplet order parameters, which can give rise to a subdominant local magnetization in the superconducting state~\cite{hillier2012nonunitary}.\\

In NC superconductors strong SOC generically yields mixed singlet-triplet pairing, so we write the symmetry-allowed TRS-breaking gap for the $E$ channel as $\hat{\Delta}_1(\mathbf{k}) = \left[ \Delta^{(1)}_s(\mathbf{k}) + \alpha \, \mathbf{d}^{(1)}(\mathbf{k})\cdot \pmb{\sigma}\right] i \sigma_y$, where $\pmb{\sigma}=(\sigma_x,\sigma_y,\sigma_z)$ are the Pauli matrices and $\alpha$ denotes the singlet-triplet mixing ratio determined by the SOC strength. On a spherical Fermi surface, the singlet and triplet components of the $E$ irrep exhibit identical nodal structures, characterized by eight point nodes along the body-diagonal directions $\mathbf{k}=(\pm 1,\pm 1,\pm 1)/\sqrt{3}$. The mixed singlet-triplet state retains these nodal positions, leading to a TRS-breaking superconducting phase with symmetry-protected point nodes on the Fermi surface [see \fig{fig:LSC}(b)]. Similarly, for the 3D irreps $T_1$ and $T_2$, the triplet gaps exhibit point nodes at the north and south poles, while the singlet gaps possess multiple nodal points across the Fermi surface.\\

Adapting the general discussion about the nodal structure of the order parameters above to the specific case of ${\rm Ag}_2{\rm Pd}_3{\rm S}$, we note, from the topology of the Fermi surface sheets of ${\rm Ag}_2{\rm Pd}_3{\rm S}$ shown in \fig{fig:band_topology}(f) that all possible TRS breaking order parameters from the $E$-channel, $T_1$-channel and $T_2$-channel will lead to nodal superconducting gap functions. Fig.~\ref{crystal}(f)), upper critical field (shown in Fig.~\ref{crystal}(i)), and transverse-field $\mu$SR data (shown in Fig.~\ref{muon}(c)), strongly suggest realization of a fully gapped superconducting ground state in ${\rm Ag}_2{\rm Pd}_3{\rm S}$ which is thus incompatible with all the symmetry-allowed nodal TRS breaking instabilities based on considerations of superconductivity in a single band.\\

The resolution lies in recognizing that ${\rm Ag}_2{\rm Pd}_3{\rm S}$ is intrinsically multiband and hosts twelve symmetry-related Pd sites within its primitive unit cell. In such systems, time-reversal symmetry breaking need not arise from momentum-space nodal structures. Instead, a loop supercurrent order~\cite{Ghosh2021loop} can develop, in which superconductivity remains fully gapped and predominantly singlet, while nontrivial phase differences between inter-site pairing amplitudes generate microscopic circulating currents within the unit cell. When applied to ${\rm Ag}_2{\rm Pd}_3{\rm S}$, this mechanism yields a uniform singlet loop-supercurrent order parameter arising from a real-space pairing potential of the form $\ket{\Delta}=\ket{1}+i\ket{2}$, where $\ket{1}$ and $\ket{2}$ denote real-space basis functions representing the singlet pairing amplitudes on each of the twelve symmetry-related Pd sites. The resulting phase structure of the order parameter is illustrated in Fig.~\ref{fig:LSC}(c). This loop-supercurrent ground state supports finite microscopic Josephson current loops within the unit cell, leading to spontaneous time-reversal symmetry breaking at the superconducting transition~\cite{Ghosh2021loop} (see the supplementary information for details).\\

\section{Conclusions and Outlook}
We establish Ag$_2$Pd$_3$S as a rare material where structural chirality, nontrivial band topology, and unconventional superconductivity coexist within a single system. Thermodynamic measurements together with muon spin spectroscopy demonstrate the emergence of bulk superconductivity below $T_C = 1.1(2)$ K accompanied by spontaneous time-reversal symmetry breaking. This observation identifies Ag$_2$Pd$_3$S as the first confirmed chiral topological semimetal to host a fully gapped superconducting state with intrinsic TRS breaking. First-principles calculations further reveal a rich hierarchy of multifold quasiparticles near the Fermi level, including two-fold Kramers-Weyl, four-fold double Kramers-Weyl, six-fold spin-1, and four-fold spin-3/2 fermions carrying distinct topological charges and producing extended Fermi-arc surface states. A symmetry analysis within the Ginzburg-Landau framework indicates that the superconducting phase is consistent with a loop-supercurrent ground state that naturally explains the coexistence of a full superconducting gap and spontaneous TRS breaking. Taken together, these results position Ag$_2$Pd$_3$S as an experimentally realized platform in which nonsymmorphic crystal symmetry and multifold band topology directly influence the nature of the superconducting state.

Chiral multifold semimetals such as Ag$_2$Pd$_3$S, therefore constitute a growing class of quantum materials that support quasiparticle excitations beyond the traditional Weyl and Dirac frameworks. Their nonsymmorphic cubic symmetry stabilizes unconventional excitations, including spin-3/2 Rarita-Schwinger-Weyl fermions at the $\Gamma$ point and double spin-1 fermions at the $R$ point, which carry large Chern numbers and generate strong Berry curvature effects~\cite{bradlyn2016beyond,chang_2017}. These degeneracies lead to unusual surface states such as extended Fermi arcs and drumhead-like dispersions~\cite{schroter2019chiral}, together with nonlinear optical and transport responses including quantized circular photogalvanic effects, circular dichroism, and enhanced anomalous Hall conductivity~\cite{tang_2017,chang_2017,schroter2019chiral}. The observation of TRS-breaking superconductivity in such a system opens a promising direction for exploring how multifold fermions interact with unconventional superconductivity. Advanced spectroscopic and phase-sensitive probes may clarify whether the superconducting condensate inherits the nontrivial topology of these quasiparticles and whether Majorana excitations emerge at surfaces or defects. In the future, our aim is to synthesize the left-handed chiral phase of Ag$_2$Pd$_3$S, which corresponds to the natural mineral coldwellite, to allow for a direct comparison between the two enantiomorphic structures. Such a study will provide a unique platform for investigating the role of crystallographic chirality in the governing of superconductivity and topological electronic properties. The simultaneous presence of structural chirality, strong spin-orbit coupling, and TRS-breaking superconductivity also suggests the possibility of intrinsic superconducting diode effects~\cite{ma2025superconducting,Nadeem2023} and chirality-induced spin-selective transport~\cite{bloom2025using}, pointing to new opportunities for superconducting spintronics and dissipationless quantum technologies.

\section{Methods}

\textbf{Sample Synthesis and Characterization}

The polycrystalline sample of Ag${_2}$Pd${_3}$S was prepared using the solid-state reaction method in an evacuated sealed quartz tube. Phase purity and crystal structure analysis were performed using a Cu-$K_{\alpha}$ (1.5406~\AA) equipped PANalytical powder X-ray diffractometer. Rietveld refinement was performed using Fullprof software~\cite{Fullprof}. Energy-dispersive X-ray (EDAX) analysis was performed using a scanning electron microscope (SEM) to assess the elemental composition.\\

\textbf{Low-temperature Bulk Measurements}

The low-temperature magnetization measurements were performed in a dilution refrigerator, which confirmed the bulk superconducting transition, and magnetic measurements were performed using a Quantum Design MPMS XL superconducting quantum interference device magnetometer (SQUID) with a reciprocating sample option (RSO) insert. Transport and specific heat measurements were performed on a Quantum Design physical property measurement system (PPMS) equipped with a dilution capable of reaching 50~mK. For measurements down to 0.5~K, an IQuantum He$^{3}$ insert was used.\\

\textbf{$\mu$SR Experiments}

Muon spin rotation/relaxation ($\mu$SR) measurements were performed in transverse-field (TF) and zero-field (ZF) configurations to determine the superconducting gap structure and spontaneous magnetic field in the superconducting state, respectively. Data were collected on the M20 beamline of TRIUMF, Center for Molecular and Materials Science (CMMS), Vancouver, Canada. The polycrystalline sample was cut into flat plates and fixed in place using Copper grease. In the TF geometry, the initial muon spins were rotated perpendicularly to the magnetic field. We used a hall probe, muMetal, and three perpendicular sets of electromagnets to achieve a true zero field at the sample position~\cite{Zerostrayfield}. The MUSRFIT software package was used for the detailed analysis of the $\mu$SR data~\cite{MuSRfit}.\\

\textbf{Electronic Band Structure Calculations}

First-principles calculations of the electronic band structure were performed using density functional theory (DFT) as implemented in the \textsc{QUANTUM ESPRESSO} suite~\cite{Giannozzi_2009,Giannozzi_2017,Giannozzi_2020}. The generalized gradient approximation (GGA) in the Perdew-Burke-Ernzerhof (PBE) formulation~\cite{Perdew_1996} was used to describe the exchange-correlation energy, together with the projector augmented-wave (PAW) pseudopotentials. A plane-wave basis with an energy cutoff of $80$~Ry and a $\Gamma$-centered Monkhorst-Pack grid of $10 \times 10 \times 10$ $k$ points was employed to sample the Brillouin zone. The calculations were carried out using experimentally determined lattice constants and atomic coordinates obtained from the Rietveld refinement.

Maximally localized Wannier functions were generated using \textsc{WANNIER90}~\cite{Pizzi2020}, starting from Ag-$4p$, Ag-$3d$, Pd-$4s$, Pd-$4p$, Pd-$4d$, and S-$3s$ trial orbitals. The resulting Wannier-based tight-binding model was subsequently used to compute the Fermi surfaces and surface spectral functions of a semi-infinite Ag$_2$Pd$_3$S slab via the iterative Green's function approach implemented in \textsc{WannierTools}~\cite{Wu2018}.\\

\section*{Acknowledgement}
RKK and DS contributed equally to this work. RPS and SKG acknowledge financial support from Anusandhan National Research Foundation (ANRF), erstwhile Science and Engineering Research Board (SERB), Government of India, via the Core Research Grant CRG/2023/000817 and the Startup Research Grant: SRG/2023/000934, respectively. RKK and ShS acknowledge the UGC, Government of India, for providing a Senior Research Fellowship. $\mu$SR measurements performed at McMaster University received support from NSERC, Canada. DS and SKG utilized the \textit{Andromeda} server at IIT Kanpur for numerical calculations.

\section*{References}

\bibliographystyle{MSP}
\bibliography{Library}

@article{CePt3Si,
  title = {Heavy Fermion Superconductivity and Magnetic Order in Noncentrosymmetric \text{CePt$_{3}$Si}},
  author = {Bauer, E. and Hilscher, G. and Michor, H. and Paul, Ch. and Scheidt, E. W. and Gribanov, A. and Seropegin, Yu. and No\"el, H. and Sigrist, M. and Rogl, P.},
  journal = {Phys. Rev. Lett.},
  volume = {92},
  issue = {2},
  pages = {027003},
  numpages = {4},
  year = {2004},
  month = {Jan},
  publisher = {American Physical Society}
}

@article{Ghosh2020review,
year = {2020},
month = {oct},
publisher = {IOP Publishing},
volume = {33},
number = {3},
pages = {033001},
author = {Sudeep Kumar Ghosh and Michael Smidman and Tian Shang and James F Annett and Adrian D Hillier and Jorge Quintanilla and Huiqiu Yuan},
title = {Recent progress on superconductors with time-reversal symmetry breaking},
journal = {J. Phys.: Condens. Matter}
}

@article{Smidman2017,
  title={Superconductivity and spin-orbit coupling in non-centrosymmetric materials: a review},
  author={Smidman, M and Salamon, MB and Yuan, HQ and Agterberg, DF},
  journal={Rep. Prog. Phys.},
  volume={80},
  number={3},
  pages={036501},
  year={2017},
  publisher={IOP Publishing}
}

@article{Shang2020time,
  title = {Time-reversal symmetry breaking in the noncentrosymmetric \text{Zr$_{3}$Ir} superconductor},
  author = {Shang, T. and Ghosh, S. K. and Zhao, J. Z. and Chang, L.-J. and Baines, C. and Lee, M. K. and Gawryluk, D. J. and Shi, M. and Medarde, M. and Quintanilla, J. and Shiroka, T.},
  journal = {Phys. Rev. B},
  volume = {102},
  issue = {2},
  pages = {020503},
  numpages = {6},
  year = {2020},
  month = {Jul},
  publisher = {American Physical Society}
}

@article{Shang2022spin,
  title={Spin-triplet superconductivity in Weyl nodal-line semimetals},
  author={Shang, Tian and Ghosh, Sudeep K and Smidman, Michael and Gawryluk, Dariusz Jakub and Baines, Christopher and Wang, An and Xie, Wu and Chen, Ye and Ajeesh, Mukkattu O and Nicklas, Michael and others},
  journal={npj Quantum Mater.},
  volume={7},
  number={1},
  pages={35},
  year={2022},
  publisher={Nature Publishing Group UK London}
}

@article{Ghosh2020b,
  title = {Quantitative theory of triplet pairing in the unconventional superconductor {LaNiGa}$_2$},
  author = {Ghosh, Sudeep Kumar and Csire, G\'abor and Whittlesea, Philip and Annett, James F. and Gradhand, Martin and \'Ujfalussy, Bal\'azs and Quintanilla, Jorge},
  journal = {Phys. Rev. B},
  volume = {101},
  issue = {10},
  pages = {100506},
  numpages = {7},
  year = {2020},
  month = {Mar},
  publisher = {American Physical Society}
}

@article{Yadav2024,
  title = {Signature of point nodal superconductivity in the Dirac semimetal \text{PdTe}},
  author = {Yadav, C. S. and Ghosh, Sudeep Kumar and Kumar, Pankaj and Thamizhavel, A. and Paulose, P. L.},
  journal = {Phys. Rev. B},
  volume = {110},
  issue = {5},
  pages = {054515},
  numpages = {7},
  year = {2024},
  month = {Aug},
  publisher = {American Physical Society}
}

@article{Nadeem2023,
  title={The superconducting diode effect},
  author={Nadeem, Muhammad and Fuhrer, Michael S and Wang, Xiaolin},
  journal={Nat. Rev. Phys.},
  volume={5},
  number={10},
  pages={558--577},
  year={2023},
  publisher={Nature Publishing Group UK London}
}

@article{ma2025superconducting,
  title={Superconducting diode effects: Mechanisms, materials and applications},
  author={Ma, Jiajun and Zhan, Ruiya and Lin, Xiao},
  journal={Adv. Phys. Res.},
  volume={4},
  number={6},
  pages={2400180},
  year={2025},
  publisher={Wiley Online Library}
}

@article{bloom2025using,
  title={Using chiral-induced spin selectivity as a tool to improve materials and processes for energy science},
  author={Bloom, Brian P and Lingenfelder, Magal{\'\i} and Naaman, Ron and Sun, Dali and Waldeck, David H},
  journal={Nat. Rev. Mater.},
  pages={1--13},
  year={2025},
  publisher={Nature Publishing Group UK London}
}

@article{Shang2018,
  title = {Time-Reversal Symmetry Breaking in {Re}-Based Superconductors},
  author = {Shang, T. and Smidman, M. and Ghosh, S. K. and Baines, C. and Chang, L. J. and Gawryluk, D. J. and Barker, J. A. T. and Singh, R. P. and Paul, D. McK. and Balakrishnan, G. and Pomjakushina, E. and Shi, M. and Medarde, M. and Hillier, A. D. and Yuan, H. Q. and Quintanilla, J. and Mesot, J. and Shiroka, T.},
  journal = {Phys. Rev. Lett.},
  volume = {121},
  issue = {25},
  pages = {257002},
  numpages = {7},
  year = {2018},
  month = {Dec},
  publisher = {American Physical Society}
}

@article{Ghosh2022Dirac,
  title = {Time-reversal symmetry breaking superconductivity in three-dimensional {D}irac semimetallic silicides},
  author = {Ghosh, Sudeep K. and Biswas, P. K. and Xu, Chunqiang and Li, B. and Zhao, J. Z. and Hillier, A. D. and Xu, Xiaofeng},
  journal = {Phys. Rev. Res.},
  volume = {4},
  issue = {1},
  pages = {L012031},
  numpages = {7},
  year = {2022},
  month = {Mar},
  publisher = {American Physical Society}
}

@article{badger2022dirac,
  title={Dirac lines and loop at the Fermi level in the time-reversal symmetry breaking superconductor \text{LaNiGa$_2$}},
  author={Badger, Jackson R. and Quan, Yundi and Staab, Matthew C. and Sumita, Shuntaro and Rossi, Antonio and Devlin, Kasey P. and Neubauer, Kelly and Shulman, Daniel S and Fettinger, James C and Klavins, Peter and  Susan, M. Kauzlarich and  Aoki,  Dai and Vishik, Inna M. and  Pickett, Warren E. and  Taufour, Valentin },
  journal={Commun. Phys},
  volume={5},
  number={1},
  pages={22},
  year={2022},
  publisher={Nature Publishing Group UK London}
}

@article{kataria2026,
  title={Observation of Time-Reversal Symmetry Breaking in the Type-{I} Superconductor {YbSb}$_2$},
  author={Kataria, Anshu and Srivastava, Shashank and Samanta, Dibyendu and Yadav, Pushpendra and Manna, Poulami and Sharma, Suhani and Mishra, Priya and Barker, Joel and Hillier, Adrian D and Agarwal, Amit and others},
  journal={arXiv preprint arXiv:2601.07460},
  year={2026}
}

@article{Sajilesh2024hfrhge,
author = {Sajilesh, K. P. and Kushwaha, R. K. and Samanta, D. and Tula, T. and Meena, P. K. and Srivastava, S. and Singh, D. and Biswas, P. K. and Kanigel, A. and Hillier, A. D. and Ghosh, S. K. and Singh, R. P.},
title = {Time-Reversal Symmetry Breaking Superconductivity in {HfRhGe}: A Noncentrosymmetric Weyl Semimetal},
journal = {Adv. Mater.},
volume = {37},
pages = {2415721},
year = {2024}
}

@article{Annett1990,
  title={Symmetry of the order parameter for high-temperature superconductivity},
  author={Annett, James F},
  journal={Adv. Phys.},
  volume={39},
  number={2},
  pages={83--126},
  year={1990},
  publisher={Taylor \& Francis}
}

@article{Ghosh2021loop,
  title={Time-reversal symmetry breaking in superconductors through loop supercurrent order},
  author={Ghosh, Sudeep Kumar and Annett, James F and Quintanilla, Jorge},
  journal={New J. Phys.},
  volume={23},
  number={8},
  pages={083018},
  year={2021},
  publisher={IOP Publishing}
}

@article{Sigrist1991,
  title={Phenomenological theory of unconventional superconductivity},
  author={Sigrist, Manfred and Ueda, Kazuo},
  journal={Rev. Mod. Phys.},
  volume={63},
  number={2},
  pages={239},
  year={1991},
  publisher={APS}
}

@article{ASOC,
  title = {Superconducting 2D System with Lifted Spin Degeneracy: Mixed Singlet-Triplet State},
  author = {Gor'kov, Lev P. and Rashba, Emmanuel I.},
  journal = {Phys. Rev. Lett.},
  volume = {87},
  issue = {3},
  pages = {037004},
  numpages = {4},
  year = {2001},
  month = {Jul},
  publisher = {American Physical Society}
}

@article{Fullprof,
title = {Recent advances in magnetic structure determination by neutron powder diffraction},
journal = {Phys. B: Condens. Matter},
volume = {192},
number = {1},
pages = {55-69},
year = {1993},
issn = {0921-4526},
author = {Juan Rodríguez-Carvajal}
}

@book{tinkham,
  title={Introduction to superconductivity},
  author={Tinkham, Michael},
  volume={1},
  year={2004},
  publisher={Courier Corporation}
}

@article{WHH1,
  title = {Temperature and Purity Dependence of the Superconducting Critical Field, \text{H$_{c2}$. II}},
  author = {Helfand, E. and Werthamer, N. R.},
  journal = {Phys. Rev.},
  volume = {147},
  issue = {1},
  pages = {288--294},
  numpages = {0},
  year = {1966},
  month = {Jul},
  publisher = {American Physical Society}
}

@article{WHH2,
  title = {Temperature and Purity Dependence of the Superconducting Critical Field, \text{H$_{c2}$. III}. Electron Spin and Spin-Orbit Effects},
  author = {Werthamer, N. R. and Helfand, E. and Hohenberg, P. C.},
  journal = {Phys. Rev.},
  volume = {147},
  issue = {1},
  pages = {295--302},
  numpages = {0},
  year = {1966},
  month = {Jul},
  publisher = {American Physical Society}
}

@article{Pauli1,
  title = {Structure and physical properties of the noncentrosymmetric superconductor \text{Mo$_{3}$Al$_{2}$C}},
  author = {Karki, A. B. and Xiong, Y. M. and Vekhter, I. and Browne, D. and Adams, P. W. and Young, D. P. and Thomas, K. R. and Chan, Julia Y. and Kim, H. and Prozorov, R.},
  journal = {Phys. Rev. B},
  volume = {82},
  issue = {6},
  pages = {064512},
  numpages = {7},
  year = {2010},
  month = {Aug},
  publisher = {American Physical Society}
}

@article{Pauli2,
  title = {Superconductivity in Quasi-One-Dimensional \text{K$_{2}$Cr$_{3}$As$_{3}$} with Significant Electron Correlations},
  author = {Bao, Jin-Ke and Liu, Ji-Yong and Ma, Cong-Wei and Meng, Zhi-Hao and Tang, Zhang-Tu and Sun, Yun-Lei and Zhai, Hui-Fei and Jiang, Hao and Bai, Hua and Feng, Chun-Mu and Xu, Zhu-An and Cao, Guang-Han},
  journal = {Phys. Rev. X},
  volume = {5},
  issue = {1},
  pages = {011013},
  numpages = {6},
  year = {2015},
  month = {Feb},
  publisher = {American Physical Society}
}

@article{lambda,
  title = {Physical Properties of the Noncentrosymmetric Superconductor \text{Mg$_{10}$Ir$_{19}$B$_{16}$}},
  author = {Klimczuk, T. and Ronning, F. and Sidorov, V. and Cava, R. J. and Thompson, J. D.},
  journal = {Phys. Rev. Lett.},
  volume = {99},
  issue = {25},
  pages = {257004},
  numpages = {4},
  year = {2007},
  month = {Dec},
  publisher = {American Physical Society}
}

@article{McM,
  title = {Transition Temperature of Strong-Coupled Superconductors},
  author = {McMillan, W. L.},
  journal = {Phys. Rev.},
  volume = {167},
  issue = {2},
  pages = {331--344},
  numpages = {0},
  year = {1968},
  month = {Mar},
  publisher = {American Physical Society}
}

@article{Zerostrayfield,
  title={A method of achieving accurate zero-field conditions using muonium},
  author={Morris, GD and Heffner, RH},
  journal={Phys. B: Condens. Matter},
  volume={326},
  number={1-4},
  pages={252--254},
  year={2003},
  publisher={Elsevier}
}

@article{EHB,
  title = {Properties of the ideal Ginzburg-Landau vortex lattice},
  author = {Brandt, Ernst Helmut},
  journal = {Phys. Rev. B},
  volume = {68},
  issue = {5},
  pages = {054506},
  numpages = {11},
  year = {2003},
  month = {Aug},
  publisher = {American Physical Society}
}

@article{KT,
  title = {Zero-and low-field spin relaxation studied by positive muons},
  author = {Hayano, R. S. and Uemura, Y. J. and Imazato, J. and Nishida, N. and Yamazaki, T. and Kubo, R.},
  journal = {Phys. Rev. B},
  volume = {20},
  issue = {3},
  pages = {850--859},
  numpages = {0},
  year = {1979},
  month = {Aug},
  publisher = {American Physical Society}
}

@article{CaPd2Ge2,
  title = {Time-reversal symmetry breaking and $s\text{\ensuremath{-}}\text{wave}$ superconductivity in \text{CaPd$_2$Ge$_2$}: A \text{$\mu$SR} study},
  author = {Anand, V. K. and Bhattacharyya, A. and Adroja, D. T. and Panda, K. and Biswas, P. K. and Hillier, A. D. and Lake, B.},
  journal = {Phys. Rev. B},
  volume = {108},
  issue = {22},
  pages = {224519},
  numpages = {7},
  year = {2023},
  month = {Dec},
  publisher = {American Physical Society},
}

@article{MuSRfit,
title = {Musrfit: A Free Platform-Independent Framework for $\mu$SR Data Analysis},
journal = {Phys. Procedia},
volume = {30},
pages = {69-73},
year = {2012},
issn = {1875-3892},
author = {A. Suter and B.M. Wojek}
}

@Article{chiral,
AUTHOR = {Fecher, Gerhard H. and K\"ubler, Jürgen and Felser, Claudia},
TITLE = {Chirality in the Solid State: Chiral Crystal Structures in Chiral and Achiral Space Groups},
JOURNAL = {Materials},
VOLUME = {15},
YEAR = {2022},
NUMBER = {17},
ARTICLE-NUMBER = {5812},
PubMedID = {36079191},
ISSN = {1996-1944},
}

@article{ZrIrSi,
  title = {Probing the superconducting ground state of \text{ZrIrSi}: A muon spin rotation and relaxation study},
  author = {Panda, K. and Bhattacharyya, A. and Adroja, D. T. and Kase, N. and Biswas, P. K. and Saha, Surabhi and Das, Tanmoy and Lees, M. R. and Hillier, A. D.},
  journal = {Phys. Rev. B},
  volume = {99},
  issue = {17},
  pages = {174513},
  numpages = {6},
  year = {2019},
  month = {May},
  publisher = {American Physical Society}
}

@article{APSold,
  title={A superconducting ternary sulfide \text{Ag$_2$Pd$_3$S}},
journal = {J. Less-Common Met.},
volume = {30},
number = {1},
pages = {167-168},
year = {1973},
issn = {0022-5088},
author = {H.R. Khan and H. Trunk and Ch.J. Raub and W.A. Fertig and A.C. Lawson}
}

@article{B20,
  title = {Unconventional superconducting pairing in a B20 multifold Weyl fermion semimetal},
  author = {Mardanya, Sougata and Kargarian, Mehdi and Verma, Rahul and Chang, Tay-Rong and Chowdhury, Sugata and Lin, Hsin and Bansil, Arun and Agarwal, Amit and Singh, Bahadur},
  journal = {Phys. Rev. Mater.},
  volume = {8},
  issue = {9},
  pages = {L091801},
  numpages = {8},
  year = {2024},
  month = {Sep},
  publisher = {American Physical Society}
}

@article{gao2022topological,
  title={Topological superconductivity in multifold fermion metals},
  author={Gao, Zhe Shen and Gao, Xue-Jian and He, Wen-Yu and Xu, Xiao Yan and Ng, TK and Law, KT},
  journal={Quantum Front.},
  volume={1},
  number={1},
  pages={3},
  year={2022},
  publisher={Springer}
}

@article{bradlyn2016beyond,
  title={Beyond Dirac and Weyl fermions: Unconventional quasiparticles in conventional crystals},
  author={Bradlyn, Barry and Cano, Jennifer and Wang, Zhijun and Vergniory, MG and Felser, C and Cava, Robert Joseph and Bernevig, B Andrei},
  journal={Science},
  volume={353},
  number={6299},
  pages={aaf5037},
  year={2016},
  publisher={American Association for the Advancement of Science}
}

@article{menichetti2025chirality,
  title={Chirality-induced spin polarization in twisted transition metal dichalcogenides},
  author={Menichetti, Guido and Cavicchi, Lorenzo and Lucchesi, Leonardo and Taddei, Fabio and Iannaccone, Giuseppe and Jarillo-Herrero, Pablo and Felser, Claudia and Koppens, Frank HL and Polini, Marco},
  journal={Newton},
  volume={1},
  number={1},
  year={2025},
  publisher={Elsevier}
}

@article{Perdew_1996,
  title = {Generalized Gradient Approximation Made Simple},
  author = {Perdew, John P. and Burke, Kieron and Ernzerhof, Matthias},
  journal = {Phys. Rev. Lett.},
  volume = {77},
  issue = {18},
  pages = {3865--3868},
  numpages = {0},
  year = {1996},
  month = {Oct},
  publisher = {American Physical Society}
}

@article{Giannozzi_2009,
year = {2009},
month = {sep},
publisher = {},
volume = {21},
number = {39},
pages = {395502},
author = {Paolo Giannozzi et al.},
title = {QUANTUM ESPRESSO: a modular and open-source software project for quantum
simulations of materials},
journal = {J. Phys. Condens. Matter}
}

@article{Giannozzi_2017,
year = {2017},
month = {oct},
publisher = {IOP Publishing},
volume = {29},
number = {46},
pages = {465901},
author = {Paolo Giannozzi et al.},
title = {Advanced capabilities for materials modelling with Quantum ESPRESSO},
journal = {J. Phys. Condens. Matter}
}

@article{Giannozzi_2020,
    author = {Giannozzi, Paolo and Baseggio, Oscar and Bonfà, Pietro and Brunato, Davide and Car, Roberto and Carnimeo, Ivan and Cavazzoni, Carlo and de Gironcoli, Stefano and Delugas, Pietro and Ferrari Ruffino, Fabrizio and Ferretti, Andrea and Marzari, Nicola and Timrov, Iurii and Urru, Andrea and Baroni, Stefano},
    title = "{Quantum ESPRESSO toward the exascale}",
    journal = {J. Chem. Phys.},
    volume = {152},
    number = {15},
    pages = {154105},
    year = {2020},
    month = {04},
    issn = {0021-9606}
}

@book{sohncke1879,
  title={Entwickelung einer Theorie der Krystallstruktur},
  author={Sohncke, Leonhard},
  year={1879},
  publisher={BG Teubner}
}

@article{multunas2023circular,
  title={Circular dichroism of crystals from first principles},
  author={Multunas, Christian and Grieder, Andrew and Xu, Junqing and Ping, Yuan and Sundararaman, Ravishankar},
  journal={Phys. Rev. Mater.},
  volume={7},
  number={12},
  pages={123801},
  year={2023},
  publisher={APS}
}

@article{ideue2017bulk,
  title={Bulk rectification effect in a polar semiconductor},
  author={Ideue, T and Hamamoto, K and Koshikawa, S and Ezawa, M and Shimizu, S and Kaneko, Y and Tokura, Y and Nagaosa, N and Iwasa, Y},
  journal={Nat. Phys.},
  volume={13},
  number={6},
  pages={578--583},
  year={2017},
  publisher={Nature Publishing Group UK London}
}

@article{chang_2017,
  title = {Unconventional Chiral Fermions and Large Topological Fermi Arcs in \text{RhSi}},
  author = {Chang, Guoqing and Xu, Su-Yang and Wieder, Benjamin J. and Sanchez, Daniel S. and Huang, Shin-Ming and Belopolski, Ilya and Chang, Tay-Rong and Zhang, Songtian and Bansil, Arun and Lin, Hsin and Hasan, M. Zahid},
  journal = {Phys. Rev. Lett.},
  volume = {119},
  issue = {20},
  pages = {206401},
  numpages = {6},
  year = {2017},
  month = {Nov},
  publisher = {American Physical Society}
}

@article{Pavan_2025,
author = {Meena, Pavan Kumar and Samanta, Dibyendu and Jangid, Sonika and Kushwaha, Roshan Kumar and Stewart, Rhea and Hillier, Adrian D. and Ghosh, Sudeep Kumar and Singh, Ravi Prakash},
title = {Superconductivity in Hourglass Dirac Chain Metals ({Ti, Hf}){IrGe}},
journal = {Adv. Sci.},
volume = {12},
number = {43},
pages = {e12434},
year = {2025}
}

@article{Pavan_2025_Zr,
  title = {Nonsymmorphic symmetry protected hourglass Dirac chain topology and conventional superconductivity in \text{ZrIrGe}},
  author = {Meena, Pavan Kumar and Samanta, Dibyendu and Srivastava, Shashank and Manna, Poulami and Ghosh, Sudeep Kumar and Singh, Ravi Prakash},
  journal = {Phys. Rev. B},
  volume = {112},
  issue = {14},
  pages = {144515},
  numpages = {9},
  year = {2025},
  month = {Oct},
  publisher = {American Physical Society}
}

@article{Nagaosa2024,
  title={Nonreciprocal transport and optical phenomena in quantum materials},
  author={Nagaosa, Naoto and Yanase, Youichi},
  journal={Annu. Rev. Condens. Matter Phys.},
  volume={15},
  number={1},
  pages={63--83},
  year={2024},
  publisher={Annual Reviews}
}

@book{bauer2012non,
  title={Non-centrosymmetric superconductors: Introduction and overview},
  author={Bauer, Ernst and Sigrist, Manfred},
  volume={847},
  year={2012},
  publisher={Springer Science \& Business Media}
}

@article{arushi2021time,
  title={Time-reversal symmetry breaking and multigap superconductivity in the noncentrosymmetric superconductor \text{La$_7$Ni$_3$}},
  author={Arushi and Singh, D and Hillier, AD and Scheurer, MS and Singh, RP},
  journal={Phys. Rev. B},
  volume={103},
  number={17},
  pages={174502},
  year={2021},
  publisher={APS}
}

@article{sudeepZrOsSi, 
AUTHOR={Ghosh, Sudeep Kumar  and Li, Bin  and Xu, Chunqiang  and Hillier, Adrian D.  and Biswas, Pabitra K.  and Xu, Xiaofeng  and Shiroka, Toni },       
TITLE={{ZrOsSi}: a {Z}$_2$ topological metal with a superconducting ground state},       
JOURNAL={Front. Phys.},       
VOLUME={11},
YEAR={2023},
pages={1256166}
}

@article{singh2014detection,
  title={Detection of time-reversal symmetry breaking in the noncentrosymmetric superconductor \text{Re$_6$Zr} using muon-spin spectroscopy},
  author={Singh, Ravi P and Hillier, Adrian D and Mazidian, B and Quintanilla, J and Annett, JF and Paul, D McK and Balakrishnan, Geetha and Lees, Martin R},
  journal={Phys. Rev. Lett.},
  volume={112},
  number={10},
  pages={107002},
  year={2014},
  publisher={APS}
}

@article{singh2017time,
  title={Time-reversal symmetry breaking in the noncentrosymmetric superconductor \text{Re$_6$Hf}: Further evidence for unconventional behavior in the $\alpha$-Mn family of materials},
  author={Singh, D and Barker, JAT and Thamizhavel, A and Paul, D McK and Hillier, AD and Singh, RP},
  journal={Phys. Rev. B},
  volume={96},
  number={18},
  pages={180501},
  year={2017},
  publisher={APS}
}

@article{singh2018time,
  title={Time-reversal symmetry breaking in the noncentrosymmetric superconductor \text{Re$_6$Ti}},
  author={Singh, D and KP, Sajilesh and Barker, JAT and Paul, D McK and Hillier, AD and Singh, RP},
  journal={Phys. Rev. B},
  volume={97},
  number={10},
  pages={100505},
  year={2018},
  publisher={APS}
}

@article{shang2020simultaneous,
  title={Simultaneous nodal superconductivity and time-reversal symmetry breaking in the noncentrosymmetric superconductor \text{CaPtAs}},
  author={Shang, Tian and Smidman, M and Wang, A and Chang, L-J and Baines, C and Lee, Min-Kai and Nie, ZY and Pang, GM and Xie, W and Jiang, WB and others},
  journal={Phys. Rev. Lett.},
  volume={124},
  number={20},
  pages={207001},
  year={2020},
  publisher={APS}
}

@article{sharma2023evidence,
  title={Evidence for nonunitary triplet-pairing superconductivity in noncentrosymmetric \text{TaRuSi} and comparison with isostructural \text{TaReSi}},
  author={Sharma, S and Kp, Sajilesh and Richards, ADS and Gautreau, J and Pula, M and Beare, J and Kojima, KM and Yoon, S and Cai, Y and Kushwaha, RK and others},
  journal={Phys. Rev. B},
  volume={108},
  number={14},
  pages={144510},
  year={2023},
  publisher={APS}
}

@article{Hillier2009,
  title={Evidence for Time-Reversal Symmetry Breaking in the Noncentrosymmetric Superconductor \text{LaNiC$_2$}},
  author={Hillier, Adrian D and Quintanilla, Jorge and Cywinski, Robert},
  journal={Phys. Rev. Lett.},
  volume={102},
  number={11},
  pages={117007},
  year={2009},
  publisher={APS}
}

@article{barker2015unconventional,
  title={Unconventional Superconductivity in \text{La$_7$Ir$_3$} Revealed by Muon Spin Relaxation: Introducing a New Family of Noncentrosymmetric Superconductor That Breaks Time-Reversal Symmetry},
  author={Barker, JAT and Singh, D and Thamizhavel, A and Hillier, AD and Lees, MR and Balakrishnan, G and Paul, D McK and Singh, RP},
  journal={Phys. Rev. Lett.},
  volume={115},
  number={26},
  pages={267001},
  year={2015},
  publisher={APS}
}

@article{singh2020time,
  title={Time-reversal-symmetry breaking and unconventional pairing in the noncentrosymmetric superconductor \text{La$_7$Rh$_3$}},
  author={Singh, D and Scheurer, MS and Hillier, AD and Adroja, DT and Singh, RP},
  journal={Phys. Rev. B},
  volume={102},
  number={13},
  pages={134511},
  year={2020},
  publisher={APS}
}

@article{srivastava2026hourglass,
  title={Hourglass Dirac chains enable intrinsic topological superconductivity in nonsymmorphic silicides},
  author={Srivastava, Shashank and Samanta, Dibyendu and Meena, Pavan Kumar and Manna, Poulami and Mishra, Priya and Sharma, Suhani and Naik, Prabin Kumar and Stewart, Rhea and Hillier, Adrian D and Ghosh, Sudeep Kumar and others},
  journal={arXiv preprint arXiv:2602.22793},
  year={2026}
}

@article{schroter2019chiral,
  title={Chiral topological semimetal with multifold band crossings and long Fermi arcs},
  author={Schr{\"o}ter, Niels BM and Pei, Ding and Vergniory, Maia G and Sun, Yan and Manna, Kaustuv and De Juan, Fernando and Krieger, Jonas A and S{\"u}ss, Vicky and Schmidt, Marcus and Dudin, Pavel and others},
  journal={Nat. Phys.},
  volume={15},
  number={8},
  pages={759--765},
  year={2019},
  publisher={Nature Publishing Group UK London}
}

@article{tang_2017,
  title = {Multiple Types of Topological Fermions in Transition Metal Silicides},
  author = {Tang, Peizhe and Zhou, Quan and Zhang, Shou-Cheng},
  journal = {Phys. Rev. Lett.},
  volume = {119},
  issue = {20},
  pages = {206402},
  numpages = {6},
  year = {2017},
  month = {Nov},
  publisher = {American Physical Society}
}

@article{shang2025discovery,
  title={Discovery of Nodal-Line Superconductivity in Chiral Crystals},
  author={Shang, Tian and Zhao, Jianzhou and Hu, Lun-Hui and Wu, Weikang and Xia, Keqi and Ajeesh, Mukkattu O and Nicklas, Michael and Xu, Yang and Zhan, Qingfeng and Gawryluk, Dariusz J and others},
  journal={Adv. Mater.},
  pages={e11385},
  year={2025},
  publisher={Wiley Online Library}
}

@article{gupta2021gap,
  title={Gap symmetry of the noncentrosymmetric superconductor \text{W$_{3}$Al$_{2}$C}},
  author={Gupta, R and Ying, TP and Qi, YP and Hosono, H and Khasanov, R},
  journal={Phys. Rev. B},
  volume={103},
  number={17},
  pages={174511},
  year={2021},
  publisher={APS}
}

@book{geron2022hands,
  title={Hands-on machine learning with Scikit-Learn, Keras, and TensorFlow},
  author={G{\'e}ron, Aur{\'e}lien},
  year={2022},
  publisher={" O'Reilly Media, Inc."}
}

@article{tula2021machine,
  title={Machine learning approach to muon spectroscopy analysis},
  author={Tula, Tymoteusz and M{\"o}ller, Gunnar and Quintanilla, Jorge and Giblin, Sean R and Hillier, Adrian D and McCabe, Emma E and Ramos, Silvia and Barker, Dylan S and Gibson, Stuart},
  journal={J. Phys.: Condens. Matter},
  volume={33},
  number={19},
  pages={194002},
  year={2021},
  publisher={IOP Publishing}
}

@article{hillier2012nonunitary,
  title={Nonunitary triplet pairing in the centrosymmetric superconductor \text{LaNiGa$_{2}$}},
  author={Hillier, Adrian D and Quintanilla, Jorge and Mazidian, Bayan and Annett, James F and Cywinski, Robert},
  journal={Phys. Rev. Lett.},
  volume={109},
  number={9},
  pages={097001},
  year={2012},
  publisher={APS}
}

@article{mcdonald2015coldwellite,
  title={Coldwellite, \text{Pd$_{3}$Ag$_{2}$S}, a new mineral species from the Marathon deposit, Coldwell Complex, Ontario, Canada},
  author={McDonald, Andrew M and Cabri, Louis J and Stanley, Chris J and Good, David J and Redpath, Jason and Lane, Geoff and Spratt, John and Ames, Doreen E},
  journal={Can. Mineral.},
  volume={53},
  number={5},
  pages={845--857},
  year={2015},
  publisher={Mineralogical Association of Canada}
}

@article{Wu2018,
  title={WannierTools: An open-source software package for novel topological materials},
  author={Wu, QuanSheng and Zhang, ShengNan and Song, Hai-Feng and Troyer, Matthias and Soluyanov, Alexey A},
  journal={Computer Physics Communications},
  volume={224},
  pages={405--416},
  year={2018},
  publisher={Elsevier}
}

@article{Pizzi2020,
	year = 2020,
	month = {jan},
	publisher = {{IOP} Publishing},
	volume = {32},
	number = {16},
	pages = {165902},
	author = {Giovanni Pizzi et al.},
	title = {Wannier90 as a community code: new features and applications},
	journal = {J. Phys.: Condens. Matter},
}

@article{Ag2Pd3SAPS,
  title = {Superconductivity in noncentrosymmetric $\mathrm{A}{\mathrm{g}}_{2}\mathrm{P}{\mathrm{d}}_{3}\mathrm{S}$},
  author = {Yoshida, H. and Okabe, H. and Matsushita, Y. and Isobe, M. and Takayama-Muromachi, E.},
  journal = {Phys. Rev. B},
  volume = {95},
  issue = {18},
  pages = {184514},
  numpages = {8},
  year = {2017},
  month = {May},
  publisher = {American Physical Society}
}

@article{SchaakJACS,
author = {Schaak, Raymond E. and Sra, Amandeep K. and Leonard, Brian M. and Cable, Robert E. and Bauer, John C. and Han, Yi-Fan and Means, Joel and Teizer, Winfried and Vasquez, Yolanda and Funck, Edward S.},
title = {Metallurgy in a Beaker:  Nanoparticle Toolkit for the Rapid Low-Temperature Solution Synthesis of Functional Multimetallic Solid-State Materials},
journal = {Journal of the American Chemical Society},
volume = {127},
number = {10},
pages = {3506-3515},
year = {2005},
doi = {10.1021/ja043335f},
note ={PMID: 15755172},
URL = {https://doi.org/10.1021/ja043335f},
eprint = {https://doi.org/10.1021/ja043335f}
}

@article{Carrington2003MgB2,
title = {Magnetic penetration depth of {MgB}$_{2}$},
journal = {Phys. C: Superconductivity},
volume = {385},
number = {1},
pages = {205-214},
year = {2003},
issn = {0921-4534},
doi = {https://doi.org/10.1016/S0921-4534(02)02319-5},
url = {https://www.sciencedirect.com/science/article/pii/S0921453402023195},
author = {A. Carrington and F. Manzano}
}

\clearpage

\onecolumngrid

\begin{center}
    {\Large \textbf{\textrm{Supplementary information to\texorpdfstring{\\}{A} "Unconventional Superconductivity in the Chiral Topological Semimetal \texorpdfstring{Ag$_2$Pd$_3$S}{Ag2Pd3S}"}}}
\end{center}

\vspace{20pt}


\setcounter{figure}{0} 

\renewcommand{\thefigure}{S\arabic{figure}}  

\setcounter{table}{0} 

\renewcommand{\thetable}{S\arabic{table}}  

\setcounter{equation}{1}


\twocolumngrid

\section*{SAMPLE CHARACTERIZATION}
The Rietveld-refined lattice parameters and the Wyckoff positions are listed in Table \ref{tblAPS}.

\begin{table}[htbp!]
\caption{Lattice parameters obtained from Rietveld Refinement of Ag$_{2}$Pd$_{3}$S.}
\label{tblAPS}
\setlength{\tabcolsep}{15pt}
\begin{center}
\begin{tabular}[b]{l c c c}
\hline 
\hline
Parameters &Ag$_{2}$Pd$_{3}$S\\
\hline
$a$ = $b$ = $c$ ($\text{\AA}$) & 7.2470(3)\\
V$_{cell}$ ($\text{\AA}^{3}$)&  380.61(2)\\
\hline
\end{tabular}
\par\medskip\footnotesize
\end{center}
\begin{center}
\setlength{\tabcolsep}{7.5pt}
\begin{tabular}[b]{l c c c c}\hline 
Atom & Wyckoff position & $x$ & $y$ & $z$\\
\hline
Pd & 12d & 0.1250 & 0.6945 & 0.5554\\
Ag & 8c & 0.0575 & 0.5575 & 0.9424 \\
S & 4a & 0.3750 & 0.8750 & 0.6250\\
\hline
\end{tabular}
\par\medskip\footnotesize
\end{center}
\end{table}

\begin{figure} [h] 
\centering
\includegraphics[width=\columnwidth,origin=b]{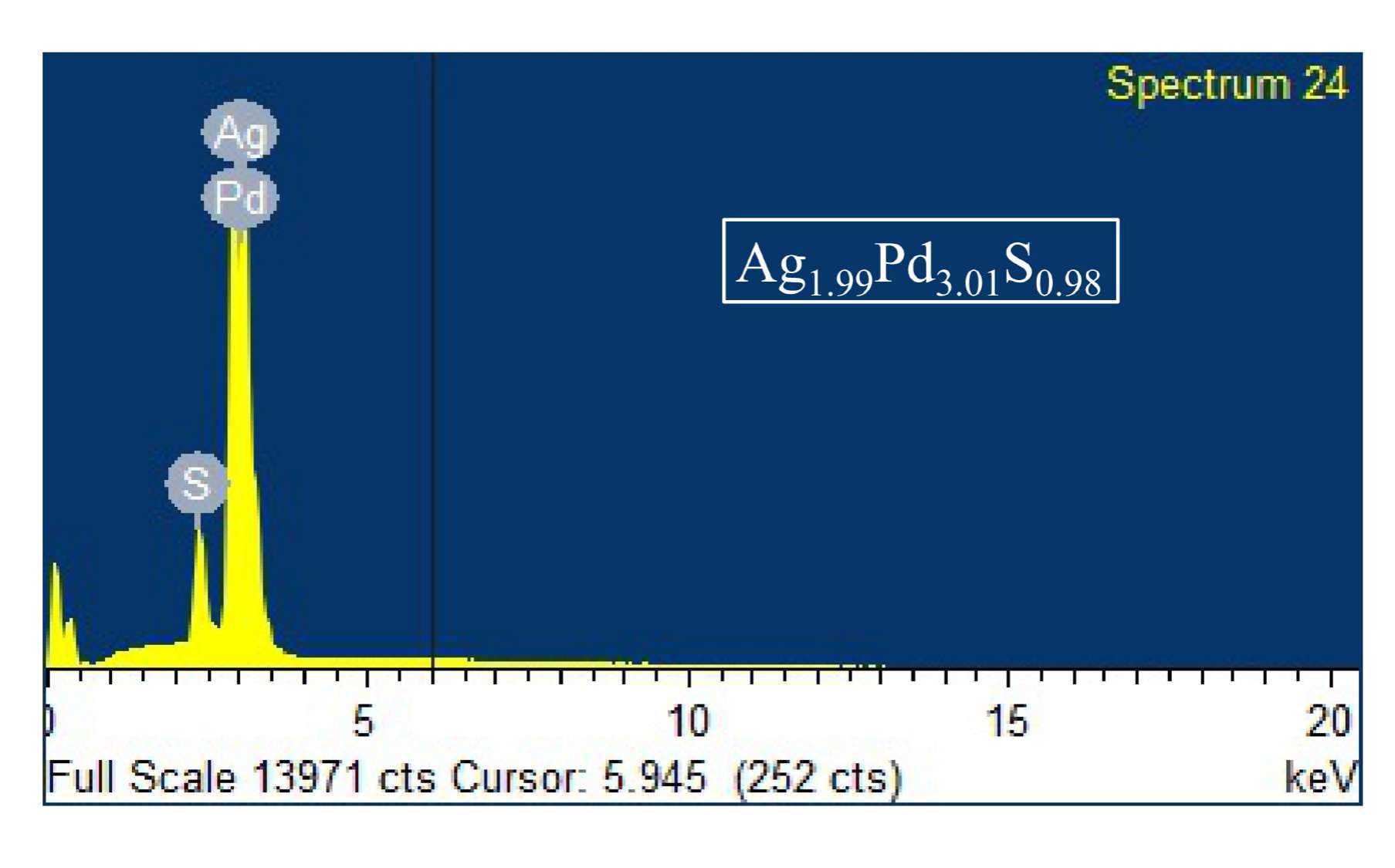}
\caption{\label{EDAX} EDAX spectrum of Ag$_{2}$Pd$_{3}$S, showing elements present in nominal composition.}
\end{figure}

The EDAX spectrum is represented in Fig. \ref{EDAX}, which confirms the elemental composition close to the nominal stoichiometry.

\section*{LOWER AND UPPER CRITICAL FIELDS}
\begin{figure} [t!] 
\centering
\includegraphics[width=0.8\columnwidth,origin=b]{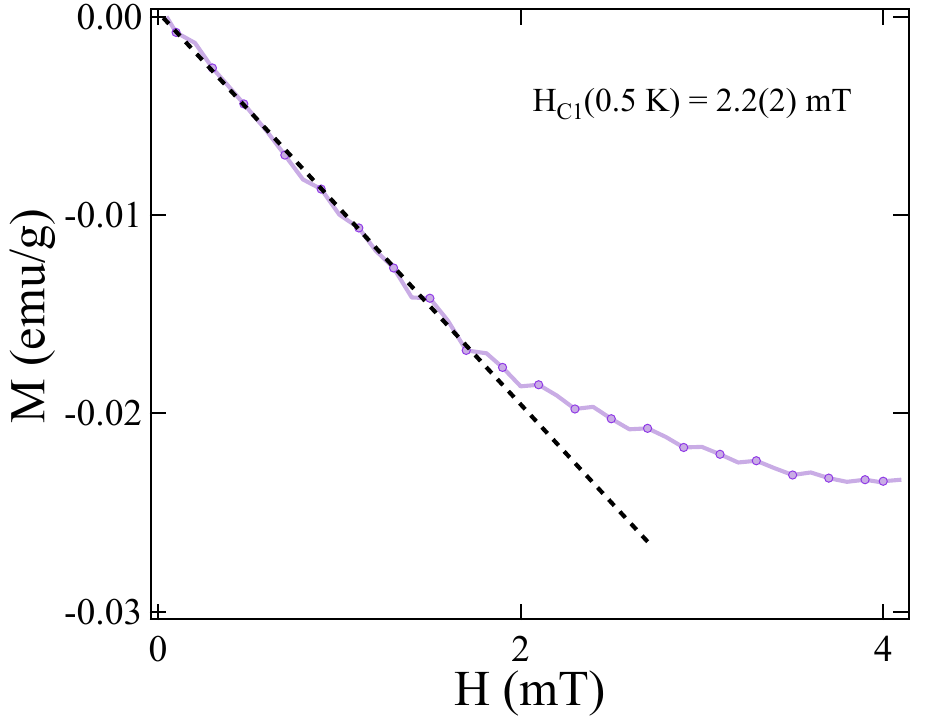}
\caption{\label{MH} Magnetization versus magnetic field plot to estimate the lower critical field.}
\end{figure}

Field-dependent magnetization (Fig.~\ref{MH}), temperature-dependent magnetization and resistivity measurements (Fig.~1(g,h), main text) were performed to evaluate the superconducting order parameters. The field-dependent magnetization curve recorded at 0.5~K shows a clear deviation from the dotted line, as shown in Fig. \ref{MH}. It estimates the lower critical field, $H_{C1}(0.5~\textrm{K})$ = 2.2(2) mT. Applying the Ginzburg-Landau (GL) relation,
\begin{equation}
  H_{C1}(T) = H_{C1}(0)\left[1 - t^{2}\right]; \quad t = T/T_{C}.
\end{equation}
$H_{C1}$(0) is extrapolated to be 2.7(1) mT. Further, the upper critical field $H_{C2}(0)$ was determined from temperature-dependent magnetization and resistivity measurements in various applied magnetic fields. The onset of the diamagnetic response in magnetization and the 50$\%$ drop of resistivity data were used to define $T_{C}$ in different magnetic fields, and $H_{C2}(0)$ was plotted as a function of temperature. Temperature-dependent $H_{C2}$ curve for Ag$_{2}$Pd$_{3}$S follows the GL equation for the upper critical field (Eq.~\ref{Hc2}).
\begin{equation}
\label{Hc2}
H_{C2}(T) = H_{C2}(0)\frac{(1-t^{2})}{(1+t^{2})}; \quad t = T/T_{C}.
\end{equation}

The GL fit yields $H_{C2}(0)$ = 168.5(3)~mT and 218.7(2)~mT from the analysis of temperature-dependent magnetization and resistivity data, respectively. In superconductors, an external magnetic field can break Cooper pairs through two main processes: orbital depairing and Pauli paramagnetic limiting. The orbital effect occurs when the field drives circulating currents, producing Lorentz forces that destroy the time-reversal symmetry required for superconductivity. The Pauli limit, on the other hand, comes from the Zeeman splitting of electron spins. The orbital limiting field is estimated by the Werthamer-Helfand-Hohenberg (WHH) model~\cite{WHH1, WHH2}, neglecting the effect of spin-orbit interaction and Pauli paramagnetism, which is described by
\begin{equation}
H_{C2}^{orb}(0)= -\alpha~T_{C} \left.\frac{dH_{C2}(T)}{dT}\right\vert_{T=T_{C}}.
\end{equation}
The initial slope at $T = T_{C}$ yields $\frac{-dH_{C2}(T)}{dT}$ = 0.16 T/K. For dirty limit superconductors, $\alpha$ = 0.693, giving an orbital limiting field, $H_{C2}^{orb}(0)$ = 125.2~mT. For BCS superconductors, the Pauli limiting field is given by $H_{C2}^{P}$= const. $\times$ $T_{C}$ where, const. = 1.86~T/K~\cite{Pauli1,Pauli2}. So, using $T_{C}$ = 1.1(2)~K, we estimate $H_{C2}^{P}$ = 2.10(2)~T for Ag$_{2}$Pd$_{3}$S, which is much higher than the estimated values of $H_{C2}(0)$.

From GL theory, once the values of $H_{C1}(0)$ and $H_{C2}(0)$ are known, the superconducting length scales can be determined, namely the coherence length and the magnetic penetration depth. The coherence length $\xi_{GL}$(0) and penetration depth $\lambda_{GL}$(0) are calculated using the following equations~\cite{lambda,Tinkham}:
\begin{equation}
\begin{split}
&H_{C2}(0)=\frac{\Phi_{0}}{2\pi\xi_{GL}^{2}(0)},\\
H_{C1}(0)=&\frac{\Phi_{0}}{4\pi\lambda_{GL}^2(0)}\left( ln \frac{\lambda_{GL}(0)}{\xi_{GL}(0)} + 0.12\right),
\end{split}
\end{equation}
where, $\Phi_{0}$ (= 2.07 $\times$10$^{-15}$ T m$^{2}$) is the magnetic flux quantum~\cite{Tinkham}. Using the estimated values of $H_{C1}$(0) = 2.7(1)~mT and $H_{C2}$(0) = 168.5(3)~mT, $\xi_{GL}$(0) and $\lambda_{GL}$(0) were evaluated to be 44.2(8)~nm and 369.9(6)~nm respectively. The GL parameter defined as $k_{GL}$ = $\frac{\lambda_{GL}(0)}{\xi_{GL}(0)}$ = 8.35(2) $>$ $\frac{1}{\sqrt{2}}$, which indicates Ag$_{2}$Pd$_{3}$S is a type-II superconductor. In addition, the thermodynamic critical field ($H_C$) is defined as 
\begin{equation}
H_{C1}(0)H_{C2}(0) = H_{C}^2ln(k_{GL}).
\end{equation}
Using estimated values of $H_{C1}(0)$ and $H_{C2}(0)$, we get $H_{C}(0)$ = 14.6(1)~mT. The Maki parameter ($\alpha_{M}$) is defined to measure the relative strength of orbital and Pauli limiting effects and is represented as $\alpha_{M} = \sqrt{2} \frac{H_{C2}^{orb}}{H_{C2}^{P}}$, which yields $\alpha_{M}$ = 0.08(4) $\ll$ 1, indicating the negligible Pauli limiting effect in Ag$_{2}$Pd$_{3}$S.\\

\section*{ANALYSIS OF GAP SYMMETRY}

\subsection*{From Specific Heat Data}
\begin{figure} [b!] 
\centering
\includegraphics[width=0.8\columnwidth,origin=b]{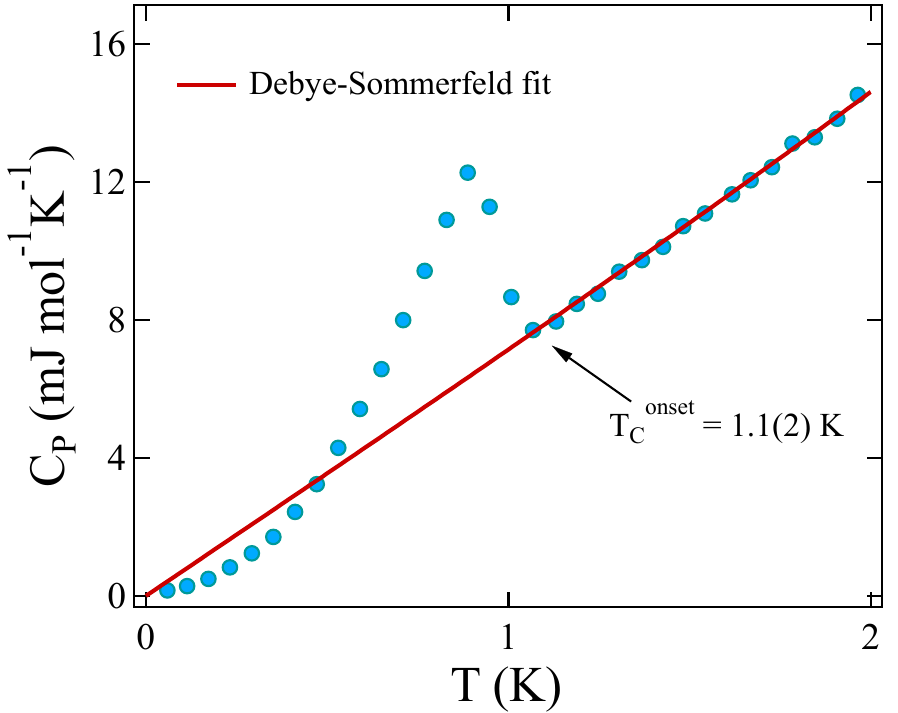}
\caption{\label{SH} Total specific heat versus temperature plot where the solid red line represents the Debye-Sommerfeld fit.}
\end{figure}

Fig. \ref{SH} represents the temperature-dependent total specific heat ($C_P$) in the absence of the applied magnetic field. The normal state region is well fitted to the Debye-Sommerfeld model described by the relation $C_P = \gamma~T + \beta~T^3$ as shown with the solid red line, where $\gamma$ represents the Sommerfeld coefficient of the normal state and $\beta$ represents the phononic contribution to the total specific heat. This fit yields $\gamma$ = 6.99(7)~mJmol$^{-1}$K$^{-2}$ and $\beta$ = 0.09(2) mJmol$^{-1}$K$^{-4}$. $\gamma$ is related to the density of states at Fermi level ($D_{C}(E_{F})$) by the relation $\gamma$ = $\left(\frac{\pi^{2}k_{B}^{2}}{3}\right)D_{C}(E_{F})$, where $k_{B}$ $\approx$ 1.38 $\times$ 10$^{-23}$~JK$^{-1}$. $D_{c}(E_{F})$ is estimated to be 2.96(6)~states/eV f.u. for Ag$_{2}$Pd$_{3}$S. The Debye temperature ($\theta_{D}$) can be obtained from $\beta$ as $\theta_{D}$ = $\left(\frac{12\pi^{4}RN}{5\beta}\right)^{1/3}$, where $N$ is the number of atoms per formula unit, $R$ is the molar gas constant (8.314~J~mol$^{-1}$K$^{-1}$), which yields $\theta_{D}$ = 506(3)~K for Ag$_{2}$Pd$_{3}$S.\\

Inverse McMillan's model~\cite{McM} estimates the electron-phonon coupling strength from a dimensionless quantity $\lambda_{e-p}$, using the estimated value of $\theta_{D}$ and $T_{C}$ as
\begin{equation}
\lambda_{e-p} = \frac{1.04 + \mu^{*}\ln(\theta_{D}/1.45T_{C})}{(1 - 0.62\mu^{*})\ln(\theta_{D}/1.45T_{C}) - 1.04},
\end{equation}
where $\mu^{*}$ represents screened Coulomb repulsion, considering $\mu^{*}$ = 0.13 in the case for all transition metals~\cite{McM}, $\theta_{D}$ = 506(3)~K and $T_{C}$ = 1.1(2)~K, we have obtained $\lambda_{e-p}$ = 0.41(4) which is close to the value of 0.47 defined for weakly coupled BCS superconductor. The electronic contribution to the total specific heat can be extracted after subtracting the phononic contribution from the total specific heat as
\begin{equation}
C_{el} = C_P - \beta~T^{3}.
\end{equation}
The temperature versus electronic specific heat curve is represented in Fig.~1(f) in the main text.\\

The symmetry of the superconducting gap can be analyzed through the temperature dependence of $C_{el}$. The relationship between the normalized entropy ($S$) in the superconducting state and $C_{el}$ is described by the following: 
\begin{equation} 
C_{el} = t \frac{dS}{dt}, \quad \text{where} ~t = \frac{T}{T_{C}}. 
\end{equation} 
In the framework of BCS theory, the normalized entropy for a single isotropic gap can be expressed as
\begin{equation}
\begin{split}
\frac{S}{\gamma T_{C}} = -\frac{6}{\pi^2} \left( \frac{\Delta(0)}{k_{B} T_{C}} \right) &\int_{0}^{\infty} f\ln(f)\\ &+ (1-f)\ln(1-f) ] dy,
\end{split}
\end{equation}
where $f = [\exp(E(\xi)/k_{B}T)+1]^{-1}$ is the Fermi function, and $E(\xi) = \sqrt{\xi^{2}+\Delta^{2}(t)}$. Here, $\xi$ denotes the energy of normal electrons relative to the Fermi energy, and $y = \xi/\Delta(0)$. The temperature variation of the superconducting gap in the BCS approximation is given by
\begin{equation}
\Delta(t) = \tanh \left[ 1.82 \left( 1.018 \left( \frac{1}{t} - 1 \right) \right)^{0.51} \right],
\end{equation}
which provides the normalized superconducting gap value $\Delta(0)/k_B T_C$ = 1.43(1), confirming weakly electron-phonon coupled superconductivity with an isotropic nodeless superconducting gap.

\subsection*{From TF-\texorpdfstring{$\mu$}{mu}SR Data}
\begin{figure} [t!] 
\centering
\includegraphics[width=0.8\columnwidth,origin=b]{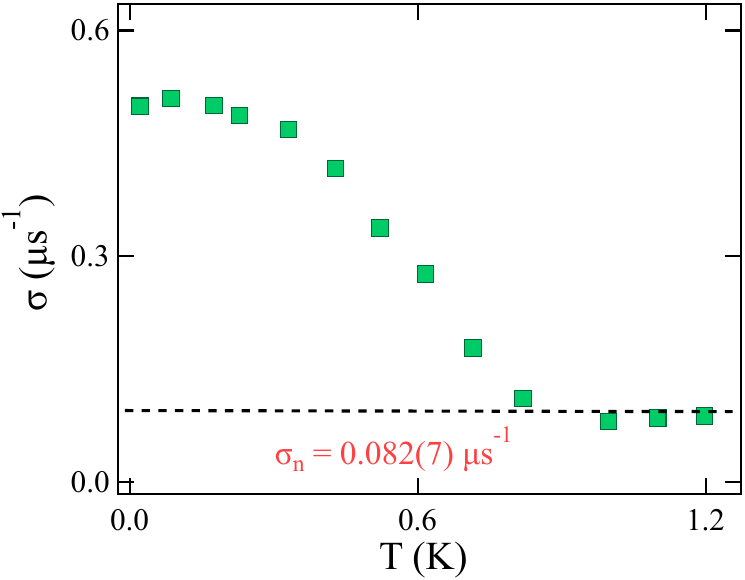}
\caption{\label{sigma} Temperature-dependent total depolarization rate, showing background nuclear contribution with the dotted line.}
\end{figure}

The temperature-dependent total depolarization rate ($\sigma$) evaluated from the analysis of TF-asymmetry data is represented in Fig. \ref{sigma}(a). Background nuclear contributions ($\sigma_n$) were extracted from Fig. \ref{sigma}(a) and the temperature-dependent superconducting contribution ($\sigma_{sc}$) was calculated, which is directly dependent on the inverse square London penetration depth ($\lambda^{-2}$) as shown in Fig. 2(c) (main text).

Following the London approximation of temperature-dependent penetration depth for a BCS superconductor,
\begin{equation}
\frac{\sigma_{sc}(T)}{\sigma_{sc}(0)} =\frac{\lambda^{-2}(T)}{\lambda^{-2}(0)} = \frac{\Delta (T)}{\Delta (0)}\tanh{\left[\frac{\Delta (T)}{2k_BT}\right]},
\label{gap}
\end{equation}
where $\Delta({T})/\Delta(0) = \tanh[1.82(1.018({T_C/T}-1))^{0.51}]$ is the BCS approximation for the temperature dependence of the energy gap \cite{Carrington2003MgB2}. The s-wave model using Eq.~\ref{gap} yields $\Delta_{1}(0)/k_{\mathrm{B}}T_{C}$ = 1.23(3). The results of the specific-heat and magnetic penetration-depth of TF-$\mu$SR consistently indicate the presence of a fully open nodeless superconducting gap in Ag$_2$Pd$_3$S.\\

\begin{figure} [b!] 
\includegraphics[width=\columnwidth,origin=b]{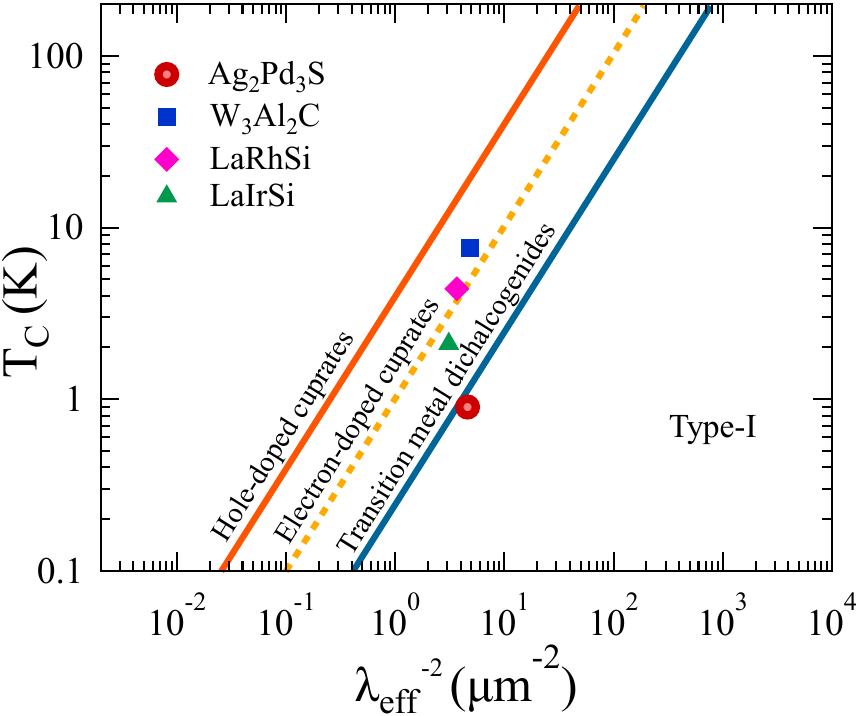}
\caption{\label{Uemura}The $T_{C}$ and $\lambda_{\mathrm{eff}}^{-2}$ plot for superconductors with chiral crystal structure \cite{gupta2021gap,shang2025discovery}.}
\end{figure}

In unconventional superconductors described by the Bose-Einstein condensation framework, the ratio $\frac{T_{C}}{\lambda_{\mathrm{eff}}^{-2}} \approx (1\text{--}20)~\mathrm{K} \mu\mathrm{m}^{2}$ exhibits a universal linear relationship between ($T_{C}$) and the superfluid stiffness ($\lambda_{\mathrm{eff}}^{-2}$)~\cite{gupta2021gap}. By contrast, conventional BCS superconductors display much smaller values, $\frac{T_{c}}{\lambda_{\mathrm{eff}}^{-2}} \approx 2.5\times10^{-4} \text{--} 1.5\times10^{-2}\;\mathrm{K}\,\mu\mathrm{m}^{2}$.
Since $\lambda_{\mathrm{eff}}^{-2}\propto n_{s}/m^{*}$, a large $T_{C}/\lambda_{\mathrm{eff}}^{-2}$ indicates an anomalously high $T_{C}$ despite a low superfluid density, hallmarks of an unconventional pairing mechanism beyond the conventional phonon-mediated BCS paradigm. For Ag$_2$Pd$_3$S, $\lambda_{\mathrm{eff}}^{-2}$ = 4.59(2)~$\mu$m$^{-2}$ and T$_C$ = 0.9(1)~K, which yields, $\frac{T_{c}}{\lambda_{\mathrm{eff}}^{-2}}$ = 0.19, which is close to the range described for unconventional superconductors. The ratio of $T_{C}$ and $\lambda_{\mathrm{eff}}^{-2}$ for some NC superconductors with chiral structure is shown in Fig. \ref{Uemura}.

\section*{PCA CALCULATIONS}
To estimate the uncertainty of the PC scores, we calculate error bars under the assumption that they originate exclusively from experimental uncertainties in the asymmetry spectra. The PC score at temperature $T_i$ is defined as
\begin{equation}
PC_{\mathrm{score}}^{[n]}(T_i) = \sum_{j=1}^{M} PC^{[n]}(t_j)\, A(T_i, t_j),
\end{equation}  
where $PC^{[n]}(t_j)$ denotes the $n$-th principal component evaluated at time $t_j$. If we treat the errors from different time bins as independent, the standard deviation of the PC score can be expressed as  
\begin{equation}
\mathrm{SD}\!\left[PC_{\mathrm{score}}^{[n]}(T_i)\right] = \sqrt{\sum_{j=1}^{M} \left( PC^{[n]}(t_j)\, E(T_i, t_j) \right)^2},
\end{equation} 
where $E(T_i, t_j)$ is the experimental uncertainty of the asymmetry $A(T_i, t_j)$. In this analysis, variations in $PC^{[n]}(t_j)$ are assumed to contribute negligibly to the overall error.

\section*{SYMMETRY ALLOWED SUPERCONDUCTING ORDER PARAMETERS IN \texorpdfstring{A\lowercase{g}$_{2}$P\lowercase{d}$_{3}$S}{Ag2Pd3S}}

Ag$_2$Pd$_3$S is a noncentrosymmetric chiral multifold semimetal that crystallizes in a primitive cubic structure. It belongs to the nonsymmorphic chiral space group $P4_132$ (No. 213), associated with the cubic point group $O$. The point group $O$ admits two one-dimensional, one two-dimensional, and two three-dimensional irreducible representations (irreps). Band structure calculations reveal that Ag$_2$Pd$_3$S exhibits strong spin-orbit coupling (SOC), which plays a crucial role in shaping its low-energy electronic states. The corresponding basis functions for the 2D and 3D irreps of the point group $O$, relevant in the presence of strong SOC, are summarized in Table \ref{tab:O}. The possible twofold and threefold degenerate superconducting instabilities associated with these 2D and 3D irreps can be constructed by expressing the superconducting order parameter as

\bea
\ket{\Delta}_{2D} = \left(\begin{array}{c} \eta_1 \\ \eta_2\end{array}\right) \quad\quad \ket{\Delta}_{3D} = \left(\begin{array}{c} \eta_1 \\ \eta_2 \\ \eta_3\end{array}\right)
\eea

\noindent where $\eta_1$, $\eta_2$, and $\eta_3$ are the complex components of the superconducting order parameter. The quartic contribution to the Ginzburg-Landau (GL) free energy is therefore expressed as
\beq
f_4 = (\bra{\Delta}{\otimes}\bra{\Delta}) \hat{\beta} (\ket{\Delta}{\otimes}\ket{\Delta}).
\eeq

{\renewcommand{\arraystretch}{1.5}{
\begin{table}[t!]
\centering
\begin{tabular}{|c|c|c|}
\hline
$O$ & \multicolumn{2}{|c|}{Basis functions} \\
\cline{2-3}
Irreps & Scalar (even) & Vector (odd) \\
\hline
$E$ & $A_1 \left(\begin{array}{c} k_x^2 - k_y^2 \\ -k_x^2 - k_y^2 + 2k_z^2 \end{array}\right)$ 
    & $A_2 \left(\begin{array}{c} k_x \hat{x} - k_y \hat{y} \\ -k_x \hat{x} - k_y \hat{y} + 2k_z \hat{z} \end{array}\right)$ \\ 
\hline
$T_1$ & $A_3 \left(\begin{array}{c} k_y k_z(k_y^2-k_z^2) \\ k_z k_x(k_z^2-k_x^2) \\ k_x k_y(k_x^2-k_y^2) \end{array}\right)$ 
      & $A_4 \left(\begin{array}{c} k_y \hat{y} - k_z \hat{z} \\ k_x \hat{z} - k_z \hat{x} \\ k_y \hat{x} - k_x \hat{y} \end{array}\right)$ \\
\hline
$T_2$ & $A_5 \left(\begin{array}{c} k_x k_y \\ k_x k_z \\ k_y k_z \end{array}\right)$ 
      & $A_6 \left(\begin{array}{c} k_y \hat{y} + k_z \hat{z} \\ k_x \hat{z} + k_z \hat{x} \\ k_y \hat{x} + k_x \hat{y} \end{array}\right)$ \\
\hline
\end{tabular}
\caption{Basis functions of the three irreducible representations of the cubic point group $O$. 
$A_i$ are constants independent of $\mathbf{k}$.}
\label{tab:O}
\end{table}
}}

\noindent For the 2D irrep $E$ of the point group $O$, the form of the tensor $\hat{\beta}$ in the GL free energy expansion is constrained by the symmetry of the normal state. The resulting quartic GL free energy, parameterized by two material-dependent coefficients $\beta$ and $\beta'$ is given by

\beq
f^{(1)}_4  =  \beta \left(|\eta_1|^2 + |\eta_2|^2\right)^2 + \beta^\prime \left(\eta^*_1 \eta_2 - \eta_1 \eta^*_2\right)^2.
\eeq

\noindent Minimizing this free energy with respect to the complex order-parameter components $\eta_1$ and $\eta_2$ yields the symmetry-allowed superconducting ground states. The minimization reveals that the possible superconducting states correspond to $(\eta_1,\eta_2) = (1,0)$ and $(\eta_1,\eta_2) = \tfrac{1}{\sqrt{2}}(1,\pm i)$. As discussed in the main text, the latter states $\tfrac{1}{\sqrt{2}}(1,\pm i)$, break time-reversal symmetry (TRS) and thus represent the TRS-breaking superconducting phases allowed by the $E$ irrep. The corresponding symmetry-allowed TRS-breaking order parameters for the $E$ representation are given by the singlet channel $\Delta^{(1)}_s(\mathbf{k}) = A_1 \left[(k_x^2 - k_y^2) - i (k_x^2 + k_y^2 - 2k_z^2)\right]$, and in the triplet channel, $\mathbf{d}^{(1)}(\mathbf{k}) = A_2 \left[(1 - i)k_x, -(1 + i)k_y, 2i k_z\right]$, where $A_1$ and $A_2$ are material-dependent constants.

The $T_1$ and $T_2$ channels each admit multiple superconducting ground states, several of which break TRS. The symmetry of the normal state constrains the GL free energy expansion in these channels. The corresponding quartic-order GL free energy, parameterized by three material-dependent coefficients $\beta$, $\beta'$, and $\beta''$, is given by

\beq
\begin{split}
f^{(2)}_4  =~&\beta \left(|\eta_1|^2 + |\eta_2|^2 + |\eta_3|^2\right)^2 + \beta^\prime |\eta_1^2 + \eta_2^2 + \eta_3^2|^2\\ &+ \beta'' (|\eta_1|^2 |\eta_2|^2 + |\eta_2|^2 |\eta_3|^2 + |\eta_3|^2 |\eta_1|^2).
\end{split}
\eeq

\noindent Minimization of the GL free energy with respect to $\eta_1$, $\eta_2$, and $\eta_3$ leads to several competing superconducting states. The symmetry-allowed TRS-breaking configuration corresponds to $\tfrac{1}{\sqrt{2}}(1, \pm i, 0)$, characteristic of complex order parameters transforming under the three-dimensional irreducible representations $T_1$ and $T_2$. The corresponding TRS-breaking superconducting order parameters for the $T_1$ irrep are given by the singlet component $\Delta^{(2)}_s(\mathbf{k}) = A_3 [k_y k_z (k_y^2 - k_z^2) + ik_z k_x (k_z^2 - k_x^2)]$ and the triplet component $\mathbf{d}^{(2)}(\mathbf{k}) = A_4 [-i k_z, k_z, (i k_x - k_y)]$. Similarly, for the $T_2$ irrep, the singlet and triplet TRS-breaking order parameters take the forms $\Delta^{(3)}_s(\mathbf{k}) = A_5 [k_x k_y + ik_x k_z]$ and $\mathbf{d}^{(3)}(\mathbf{k}) = A_6 [i k_z, k_z, (i k_x + k_y)]$, respectively.

\subsection*{Loop-supercurrent ground state}
To understand the possible superconducting instabilities in Ag$_2$Pd$_3$S, we consider only the symmetry properties of the group of $12$ symmetrically distinct Pd atoms which have the lowest symmetry. Moreover, the Pd atoms contribute most to the density of states at the Fermi level. In this case, the IPS matrix $\hat{\alpha}$ is a $12 \times 12$ real, symmetric matrix parametrized by $6$ real parameters $b_i$ ($i = 1$, $\ldots$, $6$). It takes the form
\beq\mylabel{eqn:ReAlpha}
\hat{\alpha} = \left(
\begin{array}{cccccccccccc}
 b_1 & b_2 & b_2 & b_3 & b_4 & b_5 & b_4 & b_6 & b_4 & b_4 & b_6 & b_5 \\
 b_2 & b_1 & b_3 & b_2 & b_4 & b_6 & b_4 & b_5 & b_5 & b_6 & b_4 & b_4 \\
 b_2 & b_3 & b_1 & b_2 & b_6 & b_4 & b_5 & b_4 & b_4 & b_4 & b_5 & b_6 \\
 b_3 & b_2 & b_2 & b_1 & b_5 & b_4 & b_6 & b_4 & b_6 & b_5 & b_4 & b_4 \\
 b_4 & b_4 & b_6 & b_5 & b_1 & b_2 & b_2 & b_3 & b_4 & b_5 & b_4 & b_6 \\
 b_5 & b_6 & b_4 & b_4 & b_2 & b_1 & b_3 & b_2 & b_4 & b_6 & b_4 & b_5 \\
 b_4 & b_4 & b_5 & b_6 & b_2 & b_3 & b_1 & b_2 & b_6 & b_4 & b_5 & b_4 \\
 b_6 & b_5 & b_4 & b_4 & b_3 & b_2 & b_2 & b_1 & b_5 & b_4 & b_6 & b_4 \\
 b_4 & b_5 & b_4 & b_6 & b_4 & b_4 & b_6 & b_5 & b_1 & b_2 & b_2 & b_3 \\
 b_4 & b_6 & b_4 & b_5 & b_5 & b_6 & b_4 & b_4 & b_2 & b_1 & b_3 & b_2 \\
 b_6 & b_4 & b_5 & b_4 & b_4 & b_4 & b_5 & b_6 & b_2 & b_3 & b_1 & b_2 \\
 b_5 & b_4 & b_6 & b_4 & b_6 & b_5 & b_4 & b_4 & b_3 & b_2 & b_2 & b_1 \\
\end{array}
\right).
\eeq
This IPS matrix has one non-degenerate and because of the presence of higher-dimensional irreducible representations in the crystal point group, one 2-fold degenerate and three 3-fold degenerate eigenvalues. The simplest instability which has finite loop supercurrents in Ag$_2$Pd$_3$S is associated with the doubly degenerate eigenvalue $(b_1 + 2 b_2 + b_3 - 2 b_4 - b_5 - b_6)$. The two corresponding eigenvectors $\ket{1}$ and $\ket{2}$ form an orthonormal basis in this doubly degenerate subspace. The order parameter in this degenerate subspace is written as
\beq
\ket{\Delta} = \eta_1 \ket{1} + \eta_2 \ket{2}.
\eeq

We illustrate the possibility of stabilizing the loop supercurrent state in this system by taking the parameter values $b_i = \frac{1}{2i+1}$ as an example. Then the doubly degenerate eigenvalue is $0.486136$ and the two corresponding eigenvectors are
\bea
\ket{1} &=& 
\frac{1}{2 \sqrt{6}} (-1, -1, -1, -1, -1, -1,\non \\& & -1, -1, 2, 2, 2, 2); 
\\
\ket{2} &=& 
\frac{1}{2 \sqrt{2}} (1,1,1,1,-1,-1,-1,-1,0,0,0,0)
.
\eea

The corresponding quartic order term in the Ginzburg-Landau free energy is given by
\beq
\mathcal{F}_4 = \beta \left(|\eta_1|^2 + |\eta_2|^2\right)^2 + \beta^\prime \left(\eta^*_1 \eta_2 - \eta_1 \eta^*_2\right)^2. 
\eeq
It is parametrized by the two Ginzburg-Landau parameters $\beta$ and $\beta'$. Minimizing the free energy we find that there are two possible stable ground states. The first corresponds to $(\eta_1,\eta_2) = (1,0)$ which is a conventional BCS type instability. The second is for $(\eta_1,\eta_2) =\frac{1}{\sqrt{2}}(1,i)$ which is a TRS breaking instability stabilized in the parameter regime $\frac{\beta'}{\beta} > 0$ as described in the main text. 
The ground state order parameter, for this instability, is then given by
\bea
\ket{\Delta} & = & \frac{1}{\sqrt{2}}(\ket{1} + i \ket{2}),\non \\
& = & \{\Delta_1, \Delta_1, \Delta_1, \Delta_1, \Delta_2, \Delta_2,\non \\ & & \Delta_2, \Delta_2, \Delta_3, \Delta_3, \Delta_3, \Delta_3 \} \mylabel{eqn:ins}
\eea 
where $\Delta_1 = \frac{e^{i \pi/3}}{2\sqrt{3}}$, $\Delta_2 = \frac{e^{-i \pi/3}}{2\sqrt{3}}$ and $\Delta_3 = \frac{e^{i \pi}}{2\sqrt{3}}$. Clearly, if the TRS breaking instability is realized in the two-fold degenerate channel, the superconducting ground state for Ag$_2$Pd$_3$S will have finite loop supercurrents.

\end{document}